\documentclass{llncs}

\usepackage[usenames]{color}
\usepackage{hyperref}
\usepackage{stmaryrd}
\usepackage{pslatex}
\usepackage{amssymb}
\usepackage{amsmath}
\usepackage{epsfig}
\usepackage{paralist}
\usepackage{wrapfig}
\usepackage[nofillcomment,noend,linesnumbered,noline,oldcommands]{algorithm2e}
\usepackage[version=latest]{pgf}
\usepackage{tikz}
\usepackage{color}
\usepackage{skull}
\usepackage{graphicx}
\usepackage{caption}
\usepackage{subcaption}

\usetikzlibrary{arrows,shapes,snakes,automata,positioning,backgrounds,shapes,fit,patterns,calc}

\newcommand{\todo}[1]{\textcolor{blue}{[TODO: #1]}} 
\newcommand{\rem}[1]{\textcolor{green}{[NOTE: #1]}}
\newcommand{\fixme}[1]{\textcolor{red}{[FIXME: #1]}}
\newcommand{\achtung}[1]{{\centering \textcolor{red}{!!!ACHTUNG!!!\\{\centering #1}\linebreak\hfill!!!ACHTUNG!!!\\}}}
\newcommand{\drop}[1]{\textcolor{red}{#1}}

\renewcommand{\todo}[1]{} 
\renewcommand{\rem}[1]{}
\renewcommand{\fixme}[1]{}
\renewcommand{\achtung}[1]{}
\renewcommand{\drop}[1]{}

\newcommand\figswitch[1]{#1}%

\newcommand{\entry}{e}

\newcommand{\rank}{\#}
\newcommand{\rankof}[1]{\rank{#1}}
\newcommand{\edgeof}[3]{#1\mapsto(#2,#3)}
\newcommand{\nsetof}[1]{\{1,\ldots,#1\}}
\newcommand{\fa}{F}
\newcommand{\ta}{A}
\newcommand{\forest}{f}
\newcommand{\tree}{t}
\newcommand{\graph}{g}
\newcommand{\graphprime}{\hgraph}
\newcommand{\subgraph}{h}

\newcommand{\hgraph}{h}
\newcommand{\fports}{\pi} 
\newcommand{\gports}{\phi} 
\newcommand{\rootof}[1]{\mathit{root}(#1)}

\newcommand{\lang}{L}
\newcommand{\langof}[1]{\lang(#1)}
\newcommand{\flangof}[1]{L_f(#1)}
\newcommand{\smap}{S}
\newcommand{\edge}{e}
\newcommand{\sorderof}[1]{\sqsubset_#1}

\newcommand{\trees}{t_1\cdots t_n}

\newcommand{\fport}{\fports}
\newcommand{\gport}{\gports}
\newcommand{\run}{\rho}
\newcommand{\tas}{\ta_1\cdots \ta_n}
\newcommand{\subst}[2]{[#1/#2]} 

\newcommand{\imgof}[1]{\mathit{img}(#1)}
\newcommand{\domof}[1]{\mathit{dom}(#1)}
\newcommand{\knot}{k}

\newcommand{\cpsof}[1]{\mathit{cps}(#1)}

\newcommand{\src}{s}

\newcommand{\confl}{c}

\newcommand{\reachesin}[1]{\Longrightarrow_{#1}}
\newcommand{\gsedgesof}[1]{\mathit{SE}(#1)}

\newcommand{\sedgeof}[3]{#1 \rightarrow (#2,#3)}
\newcommand{\sedge}{e}

\newcommand{\pth}{p}

\newcommand{\qspanof}[1]{\mathit{span}(#1)}
\newcommand{\unfold}{\prec}

\renewcommand{\rule}{\delta}

\newcommand{\semof}[1]{\llbracket #1 \rrbracket}
\newcommand{\nat}{\mathbb{N}}

\renewcommand{\Data}{\mathbb{D}}

\newcommand{\ptrsel}{\mathit{PSel}}
\newcommand{\eql}{\approx}
\newcommand{\datasel}{\mathit{DSel}}

\newcommand{\st}{\,|\,}
\newcommand{\abstrans}{\tau_{\code{op}}}

\newcommand{\code}[1]{\mathtt{#1}}

\newcommand{\nil}{\mathtt{null}} 

\newcommand{\stackframe}{\mathtt{sf}}

\newcommand{\length}[1]{\mathit{length}(#1)}

\newcommand{\iographof}[2]{#1_{#2}}
\newcommand{\iograph}{\iographof \graph \gports}
\newcommand{\gportsprime}{\psi}
\newcommand{\iographprime}{\iographof \graphprime \gportsprime}
\newcommand{\figtext}[1]{\small{#1}}
\newcommand{\sigof}[1]{\mathit{sig}(#1)}
\newcommand{\ithedgeof}[1]{\edge\langle#1\rangle}
\newcommand{\ithrtermof}[1]{\rule\langle#1\rangle}

\newcommand{\splitfa}{\fa_q^0}

\newcommand{\overlapof}[1]{\sim_{#1}}
\newcommand{\refinfigof}[1]{#1}

\newif\ifTR\TRfalse
\newcommand{\tronly}[1]{\ifTR{#1}\else{}\fi}

\title{Fully Automated Shape Analysis\\Based on Forest Automata
}

\author{\mbox{
  Luk\'{a}\v{s} Hol\'{\i}k \and
  Ond\v rej Leng\'al \and
  Adam Rogalewicz \and
  Ji\v{r}\'{\i} \v{S}im\'{a}\v{c}ek \and
  Tom\'{a}\v{s}~Vojnar}}
  
\institute{
  {FIT, Brno University of Technology, IT4Innovations Centre of Excellence, Czech Republic}
}

\begin{document} 

\maketitle



\begin{abstract}Forest automata (FA) have recently been proposed as a tool for
shape analysis of complex heap structures.  FA encode sets of tree
decompositions of heap graphs in the form of tuples of tree automata.  In order
to allow for representing complex heap graphs, the notion of FA allowed one to
provide user-defined FA (called boxes) that encode repetitive graph patterns of
shape graphs to be used as alphabet symbols of other, higher-level FA.  In this
paper, we propose a~novel technique of automatically learning the FA to be used
as boxes that avoids the need of providing them manually.  Further, we propose
a significant improvement of the automata abstraction used in the analysis.
The result is an efficient, fully-automated analysis that can handle even as
complex data structures as skip lists, with the performance comparable to
state-of-the-art fully-automated tools based on separation logic, which,
however, specialise in dealing with linked lists only.\end{abstract}



\section{Introduction}\label{sec:intro}

\todo{
\begin{itemize}
\item
Fixnout 4.1, blabla okolo, pridat obrazky
\item
Cist 4.1
\item
Dopsat 4.2
\item
Udelat misto (sekce 3, 5?)
\item
Cist od zacatku do konce, overovat konzistenci 
\item
Vymyslet nakou poznamku, ktera by mirne popsala, jak se lisi implementace
\item
Vsechno to naimplementovat 
\item
Overit algoritmus na prevod do interconnection uniform formy 
\item
Oddelat TODO list (az jako posledni ukol)
\end{itemize}
}

Dealing with programs that use complex dynamic linked data structures belongs
to the most challenging tasks in formal program analysis\tronly{ despite the
huge amount of research effort that has already been invested into it}. The
reason is a necessity of coping with infinite sets of reachable heap
configurations that have a form of complex graphs. Representing and
manipulating such sets in a sufficiently general, efficient, and automated way
is a notoriously difficult problem.

In \cite{forester11}, a notion of \emph{forest automata} (FA) has been proposed
for representing sets of reachable configurations of programs with complex
dynamic linked data structures. FA have a form of tuples of \emph{tree
automata} (TA) that encode sets of heap graphs decomposed into tuples of
\emph{tree components} whose leaves may refer back to the roots of the
components. In order to allow for dealing with complex heap graphs, FA may be
\emph{hierarchically nested} by using them as alphabet symbols of other,
higher-level FA. Alongside the notion of FA, a shape analysis applying FA in
the framework of \emph{abstract regular tree model checking} (ARTMC)
\cite{artmc12} has been proposed in \cite{forester11} and implemented in the
Forester tool. ARTMC accelerates the computation of sets of reachable program
configurations represented by FA by abstracting their component TA, which is
done by collapsing some of their states. The analysis was experimentally shown
to be capable of proving memory safety of quite rich classes of heap structures as well as to
be quite efficient. However, it relied on the user to provide the needed nested
FA---called \emph{boxes}---to be used as alphabet symbols of the top-level FA.


\newcommand{\figDLL}{
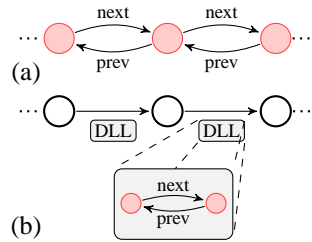
\begin{wrapfigure}[10]{r}{0.35\textwidth}
\figswitch{
  \vspace{-1mm}
  \centering
\begin{tikzpicture}[
  node distance=1.8cm,
  >=stealth',
  bend angle=45,
  auto,
  shorten >=1pt,
  shorten <=0.5pt,
  thin,
  ->,
  scale=0.8,
  every node/.style={transform shape}
]

  \tikzstyle{cell}=[circle,thick,draw=black,minimum size=5mm]
  \tikzstyle{red cell}=[circle,draw=red!75,fill=red!20,minimum size=5mm]


  \begin{scope}
	
    \node (c0) {};
    \node [red cell,right of=c0,node distance=8mm] (c1) {};
    \node [red cell,right of=c1] (c2) {};
    \node [red cell,right of=c2] (c3) {};
    \node [right of=c3,node distance=8mm](c5) {};

	\path 
         (c0) edge [dotted,thick,-] (c1)

         (c1) edge [bend left=20] node  {\figtext{next}} (c2)
         (c2) edge [bend left=20] node  {\figtext{prev}} (c1)

         (c2) edge [bend left=20] node  {\figtext{next}} (c3)
         (c3) edge [bend left=20] node  {\figtext{prev}} (c2)


         (c3) edge [dotted,thick,-] (c5)

    ;
  \end{scope}

\end{tikzpicture}\\
  \vspace{-4mm}
  \hspace*{-36mm}(a)
  \vspace{1mm}
\begin{tikzpicture}[
  node distance=1.8cm,
  >=stealth',
  bend angle=45,
  auto,
  shorten >=1pt,
  shorten <=0.5pt,
  thin,
  ->,
  scale=0.8,
  every node/.style={transform shape}
]

  \tikzstyle{cell}=[circle,thick,draw=black,minimum size=5mm]
  \tikzstyle{red cell}=[circle,draw=red!75,fill=red!20,minimum size=3mm]
  \tikzstyle{box}=[draw=black,fill=black!5,swap,yshift=-4pt, inner sep=2pt,rounded corners=1.5pt]


  \begin{scope}
	
    \node (c0) {};
    \node [cell,right of=c0,node distance=8mm] (c1) {};
    \node [cell,right of=c1] (c2) {};
    \node [cell,right of=c2] (c3) {};
    \node [right of=c3,node distance=8mm](c5) {};

	\begin{pgfonlayer}{background}
	\path 
         (c0) edge [dotted,thick,-] (c1)
         (c1) edge node [box] (b1) {\figtext{DLL}} (c2)
         (c2) edge node [box] (b2) {\figtext{DLL}} (c3)
         (c3) edge [dotted,thick,-] (c5)
    ;
	\end{pgfonlayer}

    \node [red cell, below left of=b2,node distance=1cm,xshift=-0.8cm,yshift=-0.5cm] (c6) {};
    \node [red cell, below right of=b2, node distance=1cm,xshift=-0.8cm,yshift=-0.5cm] (c7) {};

    \path
         (c6) edge [bend left=15] node [inner sep=2pt] (bla1) {\figtext{next}} (c7)
         (c7) edge [bend left=15] node [inner sep=2pt] (bla2) {\figtext{prev}} (c6)
    ;

	\begin{pgfonlayer}{background}
	\node [fit=(c6) (c7) (bla1) (bla2),rounded corners=3pt,inner sep=3pt] (big) {};

    \path
      (b2.north west) edge [dashed,-] (big.north west)
      (b2.north east) edge [dashed,-] (big.north east)
      (b2.south west) edge [dashed,-] (big.south west)
      (b2.south east) edge [dashed,-] (big.south east)
    ;

	\node [draw=black,fill=black!5,fit=(c6) (c7) (bla1) (bla2),rounded corners=3pt,inner sep=3pt] {};
	\end{pgfonlayer}

  \end{scope}

\end{tikzpicture}\\
  \vspace{-5mm}
  \hspace*{-36mm}(b)
  \vspace{-2.7mm}
}
  \caption{(a) A DLL, (b) a hierarchical encoding of a DLL.}
  \label{fig:dll}
  \vspace{-7mm}
\end{wrapfigure}
}

In this paper, we propose a new shape analysis based on FA that avoids the need
of manually providing the appropriate boxes. For that purpose, we propose a
technique of automatically \emph{learning} the FA to be used as boxes. The
basic principle of the learning  
\begin{wrapfigure}[10]{r}{0.35\textwidth}
\figswitch{
  \vspace{-1mm}
  \centering
\begin{tikzpicture}[
  node distance=1.8cm,
  >=stealth',
  bend angle=45,
  auto,
  shorten >=1pt,
  shorten <=0.5pt,
  thin,
  ->,
  scale=0.8,
  every node/.style={transform shape}
]

  \tikzstyle{cell}=[circle,thick,draw=black,minimum size=5mm]
  \tikzstyle{red cell}=[circle,draw=red!75,fill=red!20,minimum size=5mm]


  \begin{scope}
	
    \node (c0) {};
    \node [red cell,right of=c0,node distance=8mm] (c1) {};
    \node [red cell,right of=c1] (c2) {};
    \node [red cell,right of=c2] (c3) {};
    \node [right of=c3,node distance=8mm](c5) {};

	\path 
         (c0) edge [dotted,thick,-] (c1)

         (c1) edge [bend left=20] node  {\figtext{next}} (c2)
         (c2) edge [bend left=20] node  {\figtext{prev}} (c1)

         (c2) edge [bend left=20] node  {\figtext{next}} (c3)
         (c3) edge [bend left=20] node  {\figtext{prev}} (c2)


         (c3) edge [dotted,thick,-] (c5)

    ;
  \end{scope}

\end{tikzpicture}\\
  \vspace{-4mm}
  \hspace*{-36mm}(a)
  \vspace{1mm}
\begin{tikzpicture}[
  node distance=1.8cm,
  >=stealth',
  bend angle=45,
  auto,
  shorten >=1pt,
  shorten <=0.5pt,
  thin,
  ->,
  scale=0.8,
  every node/.style={transform shape}
]

  \tikzstyle{cell}=[circle,thick,draw=black,minimum size=5mm]
  \tikzstyle{red cell}=[circle,draw=red!75,fill=red!20,minimum size=3mm]
  \tikzstyle{box}=[draw=black,fill=black!5,swap,yshift=-4pt, inner sep=2pt,rounded corners=1.5pt]


  \begin{scope}
	
    \node (c0) {};
    \node [cell,right of=c0,node distance=8mm] (c1) {};
    \node [cell,right of=c1] (c2) {};
    \node [cell,right of=c2] (c3) {};
    \node [right of=c3,node distance=8mm](c5) {};

	\begin{pgfonlayer}{background}
	\path 
         (c0) edge [dotted,thick,-] (c1)
         (c1) edge node [box] (b1) {\figtext{DLL}} (c2)
         (c2) edge node [box] (b2) {\figtext{DLL}} (c3)
         (c3) edge [dotted,thick,-] (c5)
    ;
	\end{pgfonlayer}

    \node [red cell, below left of=b2,node distance=1cm,xshift=-0.8cm,yshift=-0.5cm] (c6) {};
    \node [red cell, below right of=b2, node distance=1cm,xshift=-0.8cm,yshift=-0.5cm] (c7) {};

    \path
         (c6) edge [bend left=15] node [inner sep=2pt] (bla1) {\figtext{next}} (c7)
         (c7) edge [bend left=15] node [inner sep=2pt] (bla2) {\figtext{prev}} (c6)
    ;

	\begin{pgfonlayer}{background}
	\node [fit=(c6) (c7) (bla1) (bla2),rounded corners=3pt,inner sep=3pt] (big) {};

    \path
      (b2.north west) edge [dashed,-] (big.north west)
      (b2.north east) edge [dashed,-] (big.north east)
      (b2.south west) edge [dashed,-] (big.south west)
      (b2.south east) edge [dashed,-] (big.south east)
    ;

	\node [draw=black,fill=black!5,fit=(c6) (c7) (bla1) (bla2),rounded corners=3pt,inner sep=3pt] {};
	\end{pgfonlayer}

  \end{scope}

\end{tikzpicture}\\
  \vspace{-5mm}
  \hspace*{-36mm}(b)
  \vspace{-2.7mm}
}
  \caption{(a) A DLL, (b) a hierarchical encoding of a DLL.}
  \label{fig:dll}
  \vspace{-7mm}
\end{wrapfigure}
 stems from the reason for which boxes
were originally introduced into FA. In particular, FA must have a separate
component TA for each node (called a \emph{join}) of the represented graphs
that has multiple incoming edges. If the number of joins is unbounded (as,
e.g., in doubly linked lists, abbreviated as DLLs below), unboundedly many
component TA are needed in flat FA.  However, when some of the edges are hidden
in a box (as, e.g., the prev and next links of DLLs in Fig.~\ref{fig:dll}) and
replaced by a single box-labelled edge, a finite number of component TA may
suffice. Hence, the basic idea of our learning is to identify subgraphs of the
FA-represented graphs that contain at least one join, and when they are
enclosed---or, as we say later on, \emph{folded}---into a box, the in-degree of
the join decreases. 

There are, of course, many ways to select the above mentioned subgraphs to
be used as boxes. To choose among them, we propose several criteria that we
found useful in a~number of experiments. Most importantly, the boxes must be
\emph{reusable} in order to allow eliminating as many joins as possible. The
general strategy here is to choose boxes that are \emph{simple} and
\emph{small} since these are more likely to correspond to graph patterns that
appear repeatedly in typical data structures. For instance, in the already
mentioned case of DLLs, it is enough to use a box enclosing a single pair of
next/prev links. On the other hand, as also discussed below, too simple boxes
are sometimes not useful either. 


Further, we propose a way how box learning can be efficiently integrated into
the main analysis loop. In particular, we do not use the perhaps obvious
approach of incrementally building a \emph{database of boxes} whose instances
would be sought in the generated FA. We found this approach inefficient due to
the costly operation of finding instances of different boxes in FA-represented
graphs. Instead, we always try to identify which subgraphs of the graphs
represented by a given FA could be folded into a box, followed by looking into
the so-far built database of boxes whether such a box has already been
introduced or not. Moreover, this approach has the advantage that it allows one
to use simple language inclusion checks for \emph{approximate box folding},
replacing a set of subgraphs that appear in the graphs represented by a given
FA by a larger set, which sometimes greatly accelerates the computation.
Finally, to further improve the efficiency, we interleave the process of box
learning with the \emph{automata abstraction} into a single iterative process.
In addition, we propose an FA-specific improvement of the basic automata
abstraction which \emph{accelerates the abstraction} of an FA using components
of other FA. Intuitively, it lets the abstraction synthesize an invariant
faster by allowing it to combine information coming from different branches of
the symbolic computation.

We have prototyped the proposed techniques in Forester and evaluated it on a
number of challenging case studies. The results show that the obtained approach
is both quite general as well as efficient. We were, e.g., able to
fully-automatically analyse programs with 2-level and 3-level skip lists,
which, according to the best of our knowledge, no other fully-automated
analyser can handle. On the other hand, our implementation achieves performance
comparable and sometimes even better than that of Predator \cite{predator11} (a
winner of the heap manipulation division\tronly{ of the latest software
verification competition} of SV-COMP'13) on list manipulating programs despite
being able to handle much more general classes of heap graphs.



\smallskip\noindent\emph{Related work.} As discussed already above, we propose
a new shape analysis based upon the notion of forest automata introduced in
\cite{forester11}. The new analysis is extended by a mechanism for
automatically learning the needed nested FA, which is carefully integrated into
the main analysis loop in order to maximize its efficiency. Moreover, we
formalize the abstraction used in \cite{forester11}, which was not done in
\cite{forester11}, and subsequently significantly refine it in order to improve
both its generality as well as efficiency.

From the point of view of efficiency and degree of automation, the main
alternative to our approach is the fully-automated use of separation logic with
inductive list predicates as implemented in Space Invader~\cite{InvaderCAV08} or
SLAyer~\cite{slayer11}. These approaches are, however, much less general than
our approach since they are restricted to programs over certain classes of
linked lists (and cannot handle even structures such as linked lists with data
pointers pointing either inside the list nodes or optionally outside of them,
which we can easily handle as discussed later on). A similar comparison applies
to the Predator tool inspired by separation logic but using purely graph-based
algorithms \cite{predator11}. The work \cite{overlaid11} on overlaid data
structures mentions an extension of Space Invader to trees, but this extension
is of a limited generality and requires some manual help.

In \cite{indPrSynt07}, an approach for synthesising inductive predicates in
separation logic is proposed. This approach is shown to handle even tree-like structures
with additional pointers. One of these structures, namely, the so-called mcf
trees implementing trees whose nodes have an arbitrary number of successors
linked in a DLL, is even more general than what can in principle
be described by hierarchically nested FA (to describe mcf trees, recursively
nested FA or FA based on hedge automata would be needed). On the other hand, the
approach of \cite{indPrSynt07} seems quite dependent on exploiting the fact that the encountered
data structures are built in a ``nice'' way conforming to the structure of the
predicate to be learnt (meaning, e.g., that lists are built by adding elements
at the end only), which is close to providing an inductive definition of the
data structure.

The work \cite{thor10} proposes an approach which uses separation logic for
generating numerical abstractions of heap manipulating programs allowing for
checking both their safety as well as termination. The described experiments
include even verification of programs with 2-level skip lists. However, the work
still expects the user to manually provide an inductive definition of skip lists
in advance. Likewise, the work \cite{rival11} based on the so-called separating
shape graphs reports on verification of programs with 2-level skip lists, but it
also requires the user to come up with summary edges to be used for summarizing
skip list segments, hence basically with an inductive definition of skip lists.
Compared to \cite{thor10,rival11}, we did not have to provide any manual aid
whatsoever to our technique when dealing with 2-level as well as 3-level
skip lists in our experiments.


A concept of inferring graph grammar rules for the heap abstraction proposed in
\cite{juggrnaut10} has recently appeared in \cite{juggrnaut-learning12}.
However, the proposed technique can so far only handle much less general structures
than in our case.

\section{Forest Automata}\label{sec:fa}

Given a word $\alpha = a_1\ldots a_n, n\geq 1$, we write $\alpha_i$ to denote its
$i$-th symbol $a_i$.  Given a total map $f:A\rightarrow B$, we use $\domof f$ to
denote its domain $A$ and $\imgof f$ to denote its image.


\smallskip\noindent\emph{Graphs.}
A \emph{ranked alphabet} is a finite set of symbols $\Sigma$ associated with a
mapping $\rank:\Sigma\rightarrow\nat_0$ that assigns ranks to symbols. A
(directed, ordered, labelled) \emph{graph} over $\Sigma$ is a total map
$\graph:V\rightarrow \Sigma\times V^*$ which assigns to every \emph{node} $v\in
V$ (1) a \emph{label} from $\Sigma$, denoted as $\ell_\graph(v)$, and (2) a
sequence of \emph{successors} from $V^*$, denoted as $\smap_\graph(v)$, such
that $\rankof{\ell_\graph(v)} = |\smap_\graph(v)|$.  We drop the subscript
$\graph$ if no confusion may arise.  Nodes $v$ with $\smap(v) = \epsilon$ are
called \emph{leaves}. For any $v \in V$ such that $g(v) = (a,v_1\cdots v_n)$,
we call the pair $v \mapsto (a,v_1\cdots v_n)$ an \emph{edge} of $\graph$. The
\emph{in-degree} of a node in $V$ is the overall number of its occurrences in
$\graph(v)$ across all $v \in V$. The nodes of a graph $\graph$ with an
in-degree larger than one are called \emph{joins} of $\graph$.


A \emph{path} from $v$ to $v'$ in $\graph$ is a sequence $\pth =
v_0,i_1,v_1,\ldots, i_n,v_n$ where $v_0 = v$, $v_n = v'$, and for each $j:1\leq
j\leq n$, $v_j$ is the $i_j$-th  successor of $v_{j-1}$. The \emph{length} of
$\pth$ is defined as $\length{\pth} = n$. The \emph{cost} of $\pth$ is the
sequence $i_1,\ldots,i_n$. We say that $\pth$ is cheaper than another path
$\pth'$ iff the cost of $\pth$ is lexicographically smaller than that of
$\pth'$. A node $u$
is \emph{reachable} from a node $v$ iff there is a path from $v$ to $u$ or
$u=v$. A graph $\graph$ is \emph{accessible} from a node $v$ iff all its nodes are
reachable from $v$. The node $v$ is then called the \emph{root} of $\graph$. A
\emph{tree} is a graph $\tree$ which is either empty, or it has exactly one
root and each of its nodes is the $i$-th successor of at most one node $v$ for
some $i\in\nat$. 



\smallskip\noindent\emph{Forests.}
Let $\Sigma \cap \nat = \emptyset$. A $\Sigma$-labelled \emph{forest} is a
sequence of trees  $\tree_1\cdots\tree_n$ over $(\Sigma \cup \nsetof{n})$ where
$\forall 1\leq i \leq n: \rankof i = 0$.
Leaves labelled by $i \in \nat$
are called \emph{root references}.

The forest $\trees$ represents the graph $\otimes \trees$ obtained by uniting
the trees of $\trees$, assuming w.l.o.g.~that their sets of nodes are disjoint,
and interconnecting their roots with the corresponding root references.
Formally, $\otimes \trees$ contains an edge $v\mapsto(a,v_1\cdots v_m)$ iff
there is an edge $v\mapsto(a,v'_1\cdots v'_m)$ of some tree ${t_i},1\leq i\leq
n$, s.t. for all $1\leq j\leq m$, $v_j = \rootof{\tree_k}$ if $v'_j$ is a root
reference with $\ell(v'_j) = k$, and $v_j = v'_j$ otherwise.


\smallskip\noindent\emph{Tree automata.}
A (finite, non-deterministic, top-down) \emph{tree automaton} (TA) is a
quadruple $\ta = (Q, \Sigma,\Delta, R)$ where $Q$ is a finite set of
\emph{states}, $R \subseteq Q$ is a~set of \emph{root states}, $\Sigma$ is
a~ranked alphabet, and $\Delta$ is a~set of \emph{transition rules}. Each
transition rule is a triple of the form $(q,a,q_1 \ldots q_n)$ where $n \geq
0$, $q,q_1,\ldots,q_n\in Q$, $a \in\Sigma$, and $\rankof{a} = n$. In the
special case where $n = 0$, we speak about the so-called \emph{leaf rules}.

A \emph{run} of $\ta$ over a tree $\tree$ over $\Sigma$ is a mapping $\run :
\domof\tree \rightarrow Q$ s.t. for each node $v \in \domof\tree$ where $q =
\run(v)$, if $q_i = \run(\smap(v)_i)$ for $1 \leq i \leq |\smap(v)|$, then
$\Delta$ has a~rule $q\rightarrow {\ell(v)} (q_1 \ldots q_{|\smap(v)|})$. We
write $t \reachesin \run q$ to denote that $\run$ is a run of $\ta$ over $t$
s.t. $\run(\rootof\tree)=q$. We use $t\Longrightarrow q$ to denote that
$t\reachesin \run q$ for some run $\run$. The \emph{language} of a state $q$ is
defined by $L(q)=\left\{t \mid t\Longrightarrow q \right\}$, and the
\emph{language} of $\ta$ is defined by $L(\ta)=\bigcup_{q\in R}L(q)$. 


\smallskip\noindent\emph{Graphs and forests with ports.}
We will further work with graphs with designated input and output points. An
\emph{io-graph} is a pair $(\graph, \gports)$, abbreviated as $\iograph$, where
$\graph$ is a graph and $\gports \in \domof{\graph}^+$ a sequence of
\emph{ports} in which $\gport_1$ is the \emph{input port} and $\gport_2 \cdots
\gport_{|\gports|}$ is a~sequence of \emph{output ports} such that the
occurrence of ports in $\gports$ is unique. Ports and joins of $\graph$ are
called \emph{cut-points} of $\iograph$. We use $\cpsof{\iograph}$ to denote all
cut-points of $\iograph$. We say that $\iograph$ is \emph{accessible} if it is
accessible from the input port $\gport_1$. 

An io-forest is a pair $\forest = (\trees,\fports)$ s.t. $n\geq 1$ and
$\fports\in\nsetof n^+$ is a sequence of port indices, $\fport_1$ is the
\emph{input index}, and $\fport_2\dotsc\fport_{|\fports|}$ is a sequence of
\emph{output indices}, with no repetitions of indices in $\fports$. An
io-forest encodes the io-graph $\otimes \forest$~where~the~ports of $\otimes
\trees$ are roots of the trees defined by $\fports$, i.e., $\otimes \forest =
({\otimes\trees},\rootof{\tree_{\fport_1}}\cdots\rootof{\tree_{\fport_n}})$.


\smallskip\noindent\emph{Forest automata.}
A \emph{forest automaton} (FA) over $\Sigma$ is a pair $\fa = (\tas,\fports)$
where $n\geq 1$, $\tas$ is a sequence of tree automata over $\Sigma \cup\nsetof
n$, and $\fports\in\nsetof n^+$ is a sequence of port indices as defined for
io-forests. The \emph{forest language} of $\fa$ is the set of io-forests
$\flangof{\fa} =
\langof{\ta_1}\times\cdots\times\langof{\ta_n}\times\{\fports\}$, and the
\emph{graph language} of $\fa$ is the set of io-graphs $\lang(\fa) = \{ \otimes
\forest \mid \forest \in \flangof{\fa} \}$.


\smallskip\noindent\emph{Structured labels.}
We will further work with alphabets where symbols, called \emph{structured
labels}, have an inner structure. Let $\Gamma$ be a ranked alphabet of
\emph{sub-labels}, ordered by a total ordering $\sorderof\Gamma$. We will work
with graphs over the alphabet $2^\Gamma$ where for every symbol
$A\subseteq\Gamma$, $\rankof A = \sum_{a\in A}\rankof a$. Let $\edge = \edgeof
v {\{a_1,\ldots,a_m\}} {v_1\cdots v_n}$ be an edge of a graph $\graph$ where
$n=\sum_{1\leq i\leq m}\rankof a_i$ and $a_1\sorderof\Gamma
a_2\sorderof\Gamma\cdots\sorderof\Gamma a_m$. The triple $\ithedgeof i =
\sedgeof v {a_i}  {v_k\cdots v_l}$, $1\leq i \leq m$, from the sequence
$\ithedgeof 1 = \sedgeof v {a_1}  {v_1\cdots v_{\rankof {a_1}}},\ldots,
\ithedgeof m = \sedgeof v {a_m} {v_{n-\rankof{a_{m}+1}}\cdots v_n}$ is called
the $i$-th \emph{sub-edge} of $\edge$ (or the $i$-th sub-edge of $v$ in
$\graph$). We use $\gsedgesof\graph$ to denote the set of all sub-edges of
$\graph$. We say that a node $v$ of a graph is \emph{isolated} if it does not
appear within any sub-edge, neither as an origin (i.e., $\ell(v) = \emptyset$)
nor as a target. A graph $\graph$ without isolated nodes is unambiguously
determined by $\gsedgesof\graph$ and vice versa (due to the total ordering
$\sorderof\Gamma$ and since $\graph$ has no isolated nodes). We further
restrict ourselves to graphs with structured labels and without isolated nodes. 

A counterpart of the notion of sub-edges in the context of rules of TA is the
notion of rule-terms, defined as follows: Given a rule $\rule  =
(q,\{a_1,\ldots,a_m\},q_1\cdots q_n)$ of a TA over structured labels of
$2^\Gamma$, \emph{rule-terms} of $\rule$ are the terms $\ithrtermof 1 =
a_1(q_1\cdots q_{\rankof {a_1}}),\ldots, \ithrtermof m = a_m(q_{n-\rankof
{a_{m}+1}}\cdots q_n)$ where $\ithrtermof i,1\leq i\leq m$, is called the
\emph{$i$-th rule-term of $\rule$}.  



\smallskip\noindent\emph{Forest automata of a higher level.}
We let $\Gamma_1$ be the set of all forest automata over $2^\Gamma$ and call
its elements forest automata over $\Gamma$ \emph{of level 1}. For $i > 1$, we
define $\Gamma_i$ as the set of all forest automata over ranked alphabets
$2^{\Gamma\cup \Delta}$ where $\Delta\subseteq\Gamma_{i-1}$ is any nonempty
finite set of FA of level $i-1$.  We denote elements of $\Gamma_i$ as forest
automata over $\Gamma$ \emph{of level $i$}. The rank $\rankof \fa$ of an FA
$\fa$ in these alphabets is the number of its output port indices. When used in
an FA $\fa$ over $2^{\Gamma\cup \Delta}$, the forest automata from $\Delta$ are
called \emph{boxes} of $\fa$. We write $\Gamma_*$ to denote $\cup_{i\geq
0}\Gamma_i$ and assume that $\Gamma_*$ is ordered by some total ordering
$\sorderof{{\Gamma_*}}$.

An FA $\fa$ of a higher level over $\Gamma$ accepts graphs where forest
automata of lower levels appear as sub-labels. To define the semantics of $\fa$
as a set of graphs over $\Gamma$, we need the following operation of
\emph{sub-edge replacement} where a sub-edge of a~graph is substituted by
another graph. Intuitively, the sub-edge is removed, and its origin and targets
are identified with the input and output ports of the substituted graph,
respectively.

Formally, let $\graph$ be a graph with an edge $\edge\in\graph$ and its $i$-th
sub-edge $\ithedgeof i = \sedgeof {v_1} a {v_2\cdots v_n},1\leq i\leq
|\smap_\graph(v_1)|$.  Let $\iographof {\graph'} \gports$ be an io-graph with
$|\gports|=n$. Assume w.l.o.g.~that $\domof{\graph}\cap
\domof{\graph'}=\emptyset$. The sub-edge $\ithedgeof i$ can be replaced by
$\graph'$ provided that $\forall 1 \leq j \leq n: \ell_{\graph}(v_j) \cap
\ell_{\graph'}(\gport_j) = \emptyset$, which means that the node $v_j \in
\domof\graph$ and the corresponding port $\gport_j \in \domof{\graph'}$ do not
have successors reachable over the same symbol. If the replacement can be done,
the result, denoted $\graph\subst{\iographof {\graph'} \gports}{\ithedgeof i}$,
is the graph $\graph_n$ in the sequence $\graph_0,\ldots,\graph_n$ of graphs
defined as follows: $\gsedgesof{\graph_0} =
\gsedgesof{\graph}\cup\gsedgesof{\graph'}\setminus\{\ithedgeof i\}$, and for
each $j:1\leq j\leq n$, the graph $\graph_j$ arises from $\graph_{j-1}$ by
(1)~deriving a~graph $h$ by replacing the origin of the sub-edges of the $j$-th
port $\gport_j$ of $\graph'$ by $v_j$, (2)~redirecting edges leading to
$\gport_{j}$ to $v_j$, i.e., replacing all occurrences of $\gport_j$ in $\imgof
h$ by $v_j$, and (3)~removing $\gport_j$. 

If the symbol $a$ above is an FA and $\iographof {\graph'} \gports \in L(a)$,
we say that $\hgraph = \graph\subst{\iographof {\graph'} \gports}{\ithedgeof
i}$ is an \emph{unfolding} of $\graph$, written $\graph\unfold\hgraph$.
Conversely, we say that $\graph$ arises from $\hgraph$ by \emph{folding}
$\iographof {\graph'} \gports$ into $\ithedgeof i$. Let $\unfold^*$ be the
reflexive transitive closure of $\unfold$. The \emph{$\Gamma$-semantics} of
$\graph$ is then the set of graphs $\graph'$ over $\Gamma$ s.t.
$\graph\unfold^* \graph'$, denoted $\semof\graph_\Gamma$, or just
$\semof\graph$ if no confusion may arise. For an FA $\fa$ of a higher level
over $\Gamma$, we let $\semof\fa = \bigcup_{\iograph\in\langof\fa}
(\semof\graph \times \{\gports\})$.


\smallskip\noindent\emph{Canonicity.}
We call an io-forest $\forest = (\trees,\fports)$ \emph{minimal} iff the roots
of the trees $\trees$ are the cut-points of $\otimes \forest$.
A~minimal forest representation of a graph is unique up to reordering of
$\trees$. Let the \emph{canonical ordering} of cut-points of $\otimes \forest$ be
defined by the cost of the cheapest paths leading from the input port to them.
We say that $\forest$ is \emph{canonical} iff it is minimal, $\otimes \forest$
is accessible, and the trees within $\trees$ are ordered by the canonical
ordering of their roots (which are cut-points of $\otimes\forest$). A~canonical
forest is thus a unique representation of an accessible io-graph. We say that
an FA \emph{respects canonicity} iff all forests from its forest language are
canonical. Respecting canonicity makes it possible to efficiently test FA
language inclusion by testing TA language inclusion of the respective
components of two FA. This method is precise for FA of level~$1$ and sound (not
always complete) for FA of a higher level \cite{forester11}.

In practice, we keep automata in the so called \emph{state uniform} form, which
simplifies maintaining of the canonicity respecting form \cite{forester11} (and
it is also useful when abstracting and ``folding'', as discussed in the
following). It is defined as follows. Given a node $v$ of a tree $\tree$ in an
io-forest, we define its \emph{span} as the pair $(\alpha,V)$ where
$\alpha\in\nat^*$ is the sequence of labels of root references reachable from
the root of $\tree$ ordered according to the prices of the cheapest paths to
them, and $V\subseteq\nat$ is the set of labels of references which occur more
than once in $\tree$. The state uniform form then requires that all nodes of
forests from $\langof\fa$ that are labelled by the same state $q$ in some
accepting run of $\fa$ have the same span, which we denote by $\qspanof q$.

\section{FA-based Shape Analysis}\label{sec:analysis}
\rem{This section could be probably shortened if necessary.}

We now provide a high-level overview of the main loop of our shape analysis.  
The analysis
automatically discovers memory safety errors (such as invalid dereferences of \texttt{null} or undefined pointers, double frees, or
memory leaks) and provides an~FA-represented over-approximation of the sets of
heap configurations reachable at each program line.
We consider sequential non-recursive C programs manipulating the heap. 
Each heap
cell may have several \emph{pointer selectors} and \emph{data selectors} from
some finite data domain (below, $\ptrsel$ denotes the set of pointer selectors,
$\datasel$ denotes the set of data selectors, and $\Data$ denotes the data
domain). 
\drop{
Although our implementation of the approach in the Forester tool can
handle limited pointer arithmetic and type casting, for the sake of simplicity
we do not consider these features in the following description. 
}
\drop{
The analysis
automatically discovers memory safety errors (such as invalid dereferences or
memory leaks) and provides an~FA-represented over-approximation of the sets of
heap configurations reachable at each program line.
}


\smallskip\noindent\emph{Heap representation.}
A single heap configuration is encoded as an io-graph $\iographof \graph \stackframe$ over
the ranked alphabet of structured labels $2^\Gamma$ with sub-labels from the
ranked alphabet $\Gamma = \ptrsel \cup (\datasel \times \Data)$ with the ranking
function that assigns each pointer selector 1 and each data selector 0. In this
graph, an allocated memory cell is represented by a~node $v$, and its internal
structure of selectors is given by a label $\ell_g(v) \in 2^{\Gamma}$. Values of
data selectors are stored directly in the structured label of a node as
sub-labels from $\datasel \times \Data$, so, e.g., a singly linked list cell
with the data value $42$ and the successor node $x_{\mathit{next}}$ may be
represented by a node $x$ such that $\ell_g(x) =
\{\code{next}(x_{\mathit{next}}), (\code{data}, 42)(\epsilon))\}$.  Selectors
with undefined values are represented such that the corresponding sub-labels are
not in $\ell_g(x)$. The null value is modelled as the special node $\nil$ such that
$\ell_g(\nil) = \emptyset$. The input port $\stackframe$ represents a special
node that contains the \emph{stack frame} of the analysed function, i.e.~a
structure where selectors correspond to variables of the function.

In order to represent (infinite) \emph{sets} of heap configurations, we use
state uniform FA of a higher level to represent sets of canonical io-forests
representing the heap configurations. The FA used as boxes are learnt during the
analysis using the learning algorithm presented in Sec.~\ref{sec:learning}.



\smallskip\noindent\emph{Symbolic Execution.}
The verification procedure performs standard abstract interpretation with the
abstract domain consisting of sets of state uniform FA (a single FA does not
suffice as FA are not closed under union) representing sets of heap
configurations at particular program locations. The computation starts from the
initial heap configuration given by an~FA for the io-graph $\iographof \graph \stackframe$
where $\graph$ comprises two nodes: $\nil$ and $\stackframe$ where $\ell_\graph(\stackframe) = \emptyset$.
The computation then executes abstract transformers
corresponding to program statements until the sets of FA held at program
locations stabilise. 
We note that abstract transformers corresponding to pointer manipulating statements are exact.
\drop{
For each operation $\code{op}$ in the intermediate representation of the
analysed program, the semantics of C implies existence of a function
$f_{\code{op}}$ which, when applied to the io-graph $\iographof \graph \stackframe$, gives
the io-graph $f_{\code{op}}(\iographof \graph \stackframe)$ representing the heap after
executing $\code{op}$. Based on $f_{\code{op}}$, we define for each operation
$\code{op}$ the corresponding abstract transformer $\abstrans$ with the property
that when $\abstrans$ is applied to the set of FA $\mathcal{S}$, the result is
the set of FA $\mathcal{S'} = \abstrans(\mathcal{S})$ such that $\bigcup_{F' \in
\mathcal{S'}}\semof{F'} = \{f_{\code{op}}(\iographof \graph \stackframe) \st \iographof \graph \stackframe
\in \semof{F} \wedge F \in \mathcal{S}\}$.
} 
Executing the abstract transformer
$\abstrans$ over a set of FA $\mathcal{S}$ is performed separately for
every $F \in \mathcal{S}$.
Some of boxes are first \emph{unfolded} to uncover the accessed part of the
heaps, then the update is performed. 
The detailed description of these steps can be found in~\cite{techrep}.

At junctions of program paths, the analysis computes unions of sets of FA.  At
loop points, the union is followed by widening. The widening is performed by
applying box \emph{folding} and \emph{abstraction} repeatedly in a loop on each
FA from $\mathcal{S}$ until the result stabilises. An elaboration of these two
operations, described in detail in Sec.~\ref{sec:learning}
and~\ref{sec:abstraction} respectively, belongs to the main contribution of the
presented paper.

\section{Learning of Boxes}\label{sec:learning}

Sets of graphs with an unbounded number of joins can only be described by FA
with the help of boxes. In particular, boxes allow one to replace (multiple)
incoming sub-edges of a join by a single sub-edge, and hence lower the
in-degree of the join. Decreasing the in-degree to 1 turns the join into an
ordinary node. When a~box is then used in a cycle of an FA, it effectively
generates an unbounded number of joins. 

The boxes are introduced by the operation of \emph{folding} of an FA $\fa$
which transforms $\fa$ into an FA $\fa'$ and a box $B$ used in $\fa'$ such that
$\semof\fa = \semof{\fa'}$.  However, the graphs in $\langof{\fa'}$ may contain
less joins since some of them are hidden in the box $B$, which encodes a set of
subgraphs containing a join and appearing repeatedly in the graphs of
$\langof\fa$. Before we explain folding, we give a characterisation of
subgraphs of graphs of $\langof\fa$ which we want to fold into a box $B$. Our
choice of the subgraphs to be folded is a compromise between two high-level
requirements. On the one hand, the folded subgraphs should contain incoming edges of
joins and be as simple as possible in order to be reusable. On the other hand,
the subgraphs should not be too small in order not to have to be subsequently
folded within other boxes (in the worst case, leading to generation of
unboundedly nested boxes).  Ideally, the hierarchical structuring of boxes
should respect the natural hierarchical structuring of the data structures
being handled since if this is not the case, unboundedly many boxes may again
be needed.

\subsection{Knots of Graphs}\label{sec:knots}

\newcommand{\figconfluence}{
\begin{wrapfigure}[4]{r}{27mm}
\begin{center}
\vspace{-5mm}
\figswitch{
    \vspace{-11mm}
    \hspace{-4mm}
    \begin{minipage}{27mm}
    \usetikzlibrary{arrows,backgrounds,fit,shapes}
\begin{tikzpicture}[
  node distance=5mm,
  thick,
  ->,
  >=stealth',
  level/.style={sibling distance = 12mm/#1,level distance = 7mm}
]

  \tikzstyle{vertex}=[circle,draw,inner sep=0.5mm]
  \tikzstyle{invisib}=[circle,inner sep=0.5mm]

  \node (1b) [invisib] {};
  \node (3b) [vertex,right of=1b,yshift=0mm] {};
  \node (4b) [vertex,below right of=3b,yshift=+0mm] {};
  \node (6b) [vertex,above right of=3b,yshift=-0mm] {};

  \draw [->,bend right] (3b) to (4b);
  \draw [->,bend right] (6b) to (3b);
  \draw [->,bend right,dotted] (4b) to (6b);


\end{tikzpicture}
    ~~~
    \usetikzlibrary{arrows,backgrounds,fit,shapes}
\begin{tikzpicture}[
  node distance=5mm,
  thick,
  ->,
  >=stealth',
  level/.style={sibling distance = 12mm/#1,level distance = 7mm}
]

  \tikzstyle{vertex}=[circle,draw,inner sep=0.5mm]
  \tikzstyle{invisib}=[circle,inner sep=0.5mm]

  \node (1b) [invisib] {};
  \node (3b) [vertex,below of=1b,yshift=0mm] {};
  \node (4b) [vertex,below left of=3b,yshift=+0mm] {};
  \node (6b) [vertex,below right of=3b,yshift=-0mm] {};
  \node (7b) [vertex,below of=4b,yshift=-1mm] {};
  \node (8b) [vertex,below of=6b,yshift=-1mm] {};
  \node (9b) [vertex,below of=3b,yshift=-7mm] {};

  \draw [->] (3b) to (4b);
  \draw [->] (3b) to (6b);
  \draw [->,dotted] (4b) to (7b);
  \draw [->,dotted] (6b) to (8b);
  \draw [->] (7b) to (9b);
  \draw [->] (8b) to (9b);


\end{tikzpicture}
    \end{minipage}
}
\vspace{-2mm}
        \caption{Confluence.}
  \label{fig:confluence}
\end{center}
\end{wrapfigure}
}

A graph $\subgraph$ is a \emph{subgraph} of a graph $\graph$ iff
$\gsedgesof{\graphprime} \subseteq \gsedgesof{\graph}$. The \emph{border} of
$\subgraph$ in $\graph$ is the subset of the set $\domof{\subgraph}$ of nodes
of $\subgraph$ that are incident with sub-edges in
$\gsedgesof\graph\setminus\gsedgesof\subgraph$. A~\emph{trace} from a node $u$
to a node $v$ in a graph $\graph$ is a~set of sub-edges $t =
\{\sedge_0,\ldots,\sedge_n\}\subseteq\gsedgesof\graph$ such that $n \geq 1$,
$\sedge_0$ is an outgoing sub-edge of $u$, $\sedge_n$ is an incoming sub-edge
of $v$, the origin of $\edge_i$ is one of the targets of $\edge_{i-1}$ for all
$1\leq i \leq n$, and no two sub-edges have the same origin. We call the
origins of $\edge_1, \dots, \edge_n$ the \emph{inner nodes} of the trace. A
trace from $u$ to $v$ is \tronly{
\begin{wrapfigure}[4]{r}{27mm}
\begin{center}
\vspace{-5mm}
\figswitch{
    \vspace{-11mm}
    \hspace{-4mm}
    \begin{minipage}{27mm}
    \usetikzlibrary{arrows,backgrounds,fit,shapes}
\begin{tikzpicture}[
  node distance=5mm,
  thick,
  ->,
  >=stealth',
  level/.style={sibling distance = 12mm/#1,level distance = 7mm}
]

  \tikzstyle{vertex}=[circle,draw,inner sep=0.5mm]
  \tikzstyle{invisib}=[circle,inner sep=0.5mm]

  \node (1b) [invisib] {};
  \node (3b) [vertex,right of=1b,yshift=0mm] {};
  \node (4b) [vertex,below right of=3b,yshift=+0mm] {};
  \node (6b) [vertex,above right of=3b,yshift=-0mm] {};

  \draw [->,bend right] (3b) to (4b);
  \draw [->,bend right] (6b) to (3b);
  \draw [->,bend right,dotted] (4b) to (6b);


\end{tikzpicture}
    ~~~
    \usetikzlibrary{arrows,backgrounds,fit,shapes}
\begin{tikzpicture}[
  node distance=5mm,
  thick,
  ->,
  >=stealth',
  level/.style={sibling distance = 12mm/#1,level distance = 7mm}
]

  \tikzstyle{vertex}=[circle,draw,inner sep=0.5mm]
  \tikzstyle{invisib}=[circle,inner sep=0.5mm]

  \node (1b) [invisib] {};
  \node (3b) [vertex,below of=1b,yshift=0mm] {};
  \node (4b) [vertex,below left of=3b,yshift=+0mm] {};
  \node (6b) [vertex,below right of=3b,yshift=-0mm] {};
  \node (7b) [vertex,below of=4b,yshift=-1mm] {};
  \node (8b) [vertex,below of=6b,yshift=-1mm] {};
  \node (9b) [vertex,below of=3b,yshift=-7mm] {};

  \draw [->] (3b) to (4b);
  \draw [->] (3b) to (6b);
  \draw [->,dotted] (4b) to (7b);
  \draw [->,dotted] (6b) to (8b);
  \draw [->] (7b) to (9b);
  \draw [->] (8b) to (9b);


\end{tikzpicture}
    \end{minipage}
}
\vspace{-2mm}
        \caption{Confluence.}
  \label{fig:confluence}
\end{center}
\end{wrapfigure}
} \emph{straight} iff none of
its inner nodes is a cut-point. A \emph{cycle} is a trace from a node $v$ to
$v$. A \emph{confluence} of $\iograph$ is either a cycle of $\iograph$ or it is
the union of two disjoint traces starting at a node $u$, called the
\emph{base}, and ending in the node $v$, called the \emph{tip} (for a cycle,
the base and the tip coincide)\tronly{---cf. Fig.~\ref{fig:confluence}}.

\newcommand{\figclosure}{
\begin{wrapfigure}[7]{r}{28mm}
\begin{center}
\figswitch{
    \vspace{-12mm}
    \usetikzlibrary{arrows,backgrounds,fit,shapes}
\begin{tikzpicture}[
  node distance=6.6mm,
  thick,
  ->,
  >=stealth',
  igedge/.style={red,dashed,very thick}
]
  \tikzstyle{vertex}=[circle,draw,inner sep=0.5mm]
  \tikzstyle{cutpo}=[vertex,fill]

  \node (n0n)  at (0mm,0cm) { };
  \node[cutpo,label=0:$u$] (source)  at (0mm,0cm) { };

  \node[vertex] (fork) [below left of=source] { };
  \node[vertex,above left of=fork,label=180:$x$] (source2) { };


  \node[vertex] (n22) [below right of=fork] { };
  \node[vertex] (n232) [right of=n22] { };

  \node[cutpo,label=0:$v$] (cutpoint) [below left of=n22] { };
  \node[vertex,label=0:$y$] (cutpointanex) [below of=cutpoint] { };
  \node[vertex] (cutpointattach) [left of=cutpoint] { };


  \draw (source) to[] (fork);
  \draw (fork) to[] (source2);

  \draw (fork) to[bend right] (cutpoint);

  \draw (fork) to[] (n22);
  \draw (n22) to[] (n232);
  \draw (n22) to[] (cutpoint);
  \draw (cutpoint) to[] (cutpointanex);
  \draw (cutpoint) to[bend right] (cutpointattach);
  \draw (cutpointattach) to[bend right] (cutpoint);


  \begin{pgfonlayer}{background}



    \node[draw,ellipse,fill=red!30,draw=black,label=-10:$h$,rounded corners=0mm,inner sep=0pt,fit= (fork.south) (cutpoint) (cutpointattach) (n232.west)] (r3) {};
  \end{pgfonlayer}

\end{tikzpicture}
}
\vspace{-8mm}
	\caption{Closure.}
  \label{fig:closure}
\end{center}
\end{wrapfigure}
}

\newcommand{\figUconfluence}{
\begin{wrapfigure}[6]{r}{28mm}
\vspace{-11mm}
\begin{center}
\figswitch{
    \usetikzlibrary{arrows,backgrounds,fit,shapes}
\begin{tikzpicture}[
  node distance=6.6mm,
  thick,
  ->,
  >=stealth',
  igedge/.style={red,dashed,very thick}
]
  \tikzstyle{vertex}=[circle,draw,inner sep=0.5mm]
  \tikzstyle{cutpo}=[vertex,fill]


  \node[cutpo] (fork) [] { };


  \node[vertex] (n232) [right of=fork] { };

  \node[cutpo] (cutpoint) [below of=fork] { };




  \draw (n232) to[bend left] (fork);
  \draw (n232) to[bend right] (fork);
  \draw (cutpoint) to[bend right] (n232);


  \begin{pgfonlayer}{background}



    \node[draw,ellipse,fill=blue!30,draw=black,label=180:$U$,rounded corners=0mm,inner sep=1pt,fit= (fork) (cutpoint)] (r3) {};
  \end{pgfonlayer}

\end{tikzpicture}
}
\vspace{-2mm}
	\caption{Confluence of a set of cutpoints.}
  \label{fig:Uconfluence}
\end{center}
\end{wrapfigure}
}

Given an io-graph $\iograph$, the \emph{signature} of a sub-graph $\subgraph$
of $\graph$ is the minimum subset $\sigof\subgraph$ of $\cpsof\iograph$ that
(1) contains $\cpsof\iograph\cap\domof\subgraph$ and (2) all nodes of
$\subgraph$, except 
\begin{wrapfigure}[7]{r}{28mm}
\begin{center}
\figswitch{
    \vspace{-12mm}
    \usetikzlibrary{arrows,backgrounds,fit,shapes}
\begin{tikzpicture}[
  node distance=6.6mm,
  thick,
  ->,
  >=stealth',
  igedge/.style={red,dashed,very thick}
]
  \tikzstyle{vertex}=[circle,draw,inner sep=0.5mm]
  \tikzstyle{cutpo}=[vertex,fill]

  \node (n0n)  at (0mm,0cm) { };
  \node[cutpo,label=0:$u$] (source)  at (0mm,0cm) { };

  \node[vertex] (fork) [below left of=source] { };
  \node[vertex,above left of=fork,label=180:$x$] (source2) { };


  \node[vertex] (n22) [below right of=fork] { };
  \node[vertex] (n232) [right of=n22] { };

  \node[cutpo,label=0:$v$] (cutpoint) [below left of=n22] { };
  \node[vertex,label=0:$y$] (cutpointanex) [below of=cutpoint] { };
  \node[vertex] (cutpointattach) [left of=cutpoint] { };


  \draw (source) to[] (fork);
  \draw (fork) to[] (source2);

  \draw (fork) to[bend right] (cutpoint);

  \draw (fork) to[] (n22);
  \draw (n22) to[] (n232);
  \draw (n22) to[] (cutpoint);
  \draw (cutpoint) to[] (cutpointanex);
  \draw (cutpoint) to[bend right] (cutpointattach);
  \draw (cutpointattach) to[bend right] (cutpoint);


  \begin{pgfonlayer}{background}



    \node[draw,ellipse,fill=red!30,draw=black,label=-10:$h$,rounded corners=0mm,inner sep=0pt,fit= (fork.south) (cutpoint) (cutpointattach) (n232.west)] (r3) {};
  \end{pgfonlayer}

\end{tikzpicture}
}
\vspace{-8mm}
	\caption{Closure.}
  \label{fig:closure}
\end{center}
\end{wrapfigure}
  the nodes of $\sigof\subgraph$ themselves, are
reachable by straight traces from $\sigof\subgraph$. Intuitively,
$\sigof\subgraph$ contains all cut-points of $\subgraph$ plus the closest
cut-points to $\subgraph$ which lie outside of $\subgraph$ but which are needed
so that all nodes of $\subgraph$ are reachable from the signature.  Consider
the example of the graph $g_u$ in Fig.~\ref{fig:closure} in which cut-points
are represented by $\bullet$. The signature of $g_u$ is the set $\{u, v\}$. The
signature of the highlighted subgraph $h$ is also equal to $\{u,v\}$. Given a
set $U\subseteq\cpsof\iograph$, a \emph{confluence of $U$} is a confluence of
$\iograph$ with the signature within $U$. Intuitively, the confluence of a set
of cut-points $U$ is a confluence whose cut-points belong to $U$ plus in case
the base is not a cut-point, then \tronly{
\begin{wrapfigure}[6]{r}{28mm}
\vspace{-11mm}
\begin{center}
\figswitch{
    \usetikzlibrary{arrows,backgrounds,fit,shapes}
\begin{tikzpicture}[
  node distance=6.6mm,
  thick,
  ->,
  >=stealth',
  igedge/.style={red,dashed,very thick}
]
  \tikzstyle{vertex}=[circle,draw,inner sep=0.5mm]
  \tikzstyle{cutpo}=[vertex,fill]


  \node[cutpo] (fork) [] { };


  \node[vertex] (n232) [right of=fork] { };

  \node[cutpo] (cutpoint) [below of=fork] { };




  \draw (n232) to[bend left] (fork);
  \draw (n232) to[bend right] (fork);
  \draw (cutpoint) to[bend right] (n232);


  \begin{pgfonlayer}{background}



    \node[draw,ellipse,fill=blue!30,draw=black,label=180:$U$,rounded corners=0mm,inner sep=1pt,fit= (fork) (cutpoint)] (r3) {};
  \end{pgfonlayer}

\end{tikzpicture}
}
\vspace{-2mm}
	\caption{Confluence of a set of cutpoints.}
  \label{fig:Uconfluence}
\end{center}
\end{wrapfigure}


\noindent }the closest cut-point from which the base is reachable is also from
$U$\tronly{ (cf. Fig.~\ref{fig:Uconfluence})}. Finally, the \emph{closure} of
$U$ is the smallest subgraph $\subgraph$ of $\iograph$ that (1) contains all
confluences of $U$ and (2) for every inner node $v$ of a straight trace of
$\subgraph$, it contains all straight traces from $v$ to leaves of $\graph$.
The closure of the signature $\{ u, v \}$ of the graph $g_u$ in
Fig.~\ref{fig:closure} is the highlighted subgraph $h$. Intuitively, Point 1
includes into the closure all nodes and sub-edges that appear on straight
traces between nodes of $U$ apart from those that do not lie on any confluence
(such as node $u$ in Fig.~\ref{fig:closure}). Note that nodes $x$ and $y$ in
Fig.~\ref{fig:closure}, which are leaves of $g_u$, are not in the closure as
they are not reachable from an inner node of any straight trace of $h$. The
\emph{closure of a~subgraph} $\subgraph$ of $\iograph$ is the closure of its
signature, and $\subgraph$ is \emph{closed} iff it equals its closure.


\smallskip\noindent\emph{Knots.}
For the rest of Sec.~\ref{sec:knots}, let us fix an io-graph
$\iograph\in\langof\fa$. We now introduce the notion of a knot which summarises
the desired properties of a~subgraph $\knot$ of $\graph$ that is to be folded
into a box. A \emph{knot} $\knot$ of $\iograph$ is a subgraph of $\graph$ such
that: \begin{inparaenum}[(1)] \item \label{enum:knot_is_singular_fork} $\knot$
is a~confluence, \item \label{enum:knot_union} $\knot$ is the union of two
knots with intersecting sets of sub-edges, or \item \label{enum:knot_closure}
$\knot$ is the closure of a knot. \end{inparaenum}
A \emph{decomposition} of
a~knot $\knot$ is a set of knots such that the union of their sub-edges equals
$\gsedgesof\knot$. The \emph{complexity of a decomposition} of $\knot$ is the
maximum of sizes of signatures of its elements. We define the \emph{complexity of
a knot} as the minimum of the complexities of its decompositions. 
A knot $\knot$ of complexity $n$ is
an \emph{optimal knot of complexity $n$} if it is
maximal among knots of complexity $n$ and if it has a root. The root must be reachable from the input port
of $\iograph$ by a trace that does not intersect with sub-edges of the optimal
knot. Notice that the requirement of maximality implies that optimal knots are closed.

The following lemma, proven in \cite{techrep}, implies that optimal knots are
uniquely identified by their signatures, which is crucial for the folding
algorithm presented later.

\begin{lemma} \label{lemma:closure} The signature of an optimal knot of
$\iograph$ equals the signature of its closure.\end{lemma}

\noindent Next, we explain what is the motivation behind the notion of an
optimal knot:

\emph{Confluences.} As mentioned above, in order to allow one to eliminate a
join, a knot must contain some join $v$ together with at least one incoming
sub-edge in case the knot is based on a loop and at least two sub-edges
otherwise. Since $\iograph$ is accessible (meaning that there do not exist any
traces that cannot be extended to start from the same node), the edge must
belong to some confluence $\confl$ of $\iograph$. If the folding operation does
not fold the entire $\confl$, then a new join is created on the border of the
introduced box: one of its incoming sub-edges is labelled by the box that
replaces the folded knot, another one is the last edge of one of the traces of
$\confl$. Confluences are therefore the smallest subgraphs that can be folded
in a meaningful way.

\begin{wrapfigure}[5]{r}{27mm}
\begin{center}
\vspace{-11mm}
\figswitch{
    \usetikzlibrary{arrows,backgrounds,fit}
\begin{tikzpicture}[
  node distance=6.6mm,
  thick,
  >=stealth'
]
  \tikzstyle{vertex}=[circle,draw,inner sep=0.5mm]

  \node[vertex] (n1)  at (0mm,0cm) { };
  \node[vertex] (n2) [right of=n1] { };
  \node[vertex] (n3) [right of=n2] { };
  \node[vertex] (n4) [right of=n3] { };

  \draw[->] (n1) -- (n2);
  \draw[->] (n2) -- (n3);
  \draw[->] (n3) -- (n4);

  \draw[->] (n4) to[bend right] coordinate (c1) (n1);
  \draw[->] (n3) to[bend right] coordinate (c2) (n1);
  \draw[->] (n2) to[bend right] coordinate (c3) (n1);

  \begin{pgfonlayer}{background}

    \node[draw,rectangle,fill=blue!40,draw=black,rounded corners=2mm,inner sep=5.5pt,fit= (n4) (n1) (c1)] (r1) {};
    \node[draw,rectangle,fill=red!20,draw=black,rounded corners=2mm,inner sep=4pt,fit= (n1) (n3) (c2)] (r2) {};
    \node[draw,rectangle,fill=blue!30,draw=black,rounded corners=2mm,inner sep=2pt,fit= (n1) (n2) (c3)] (r3) {};

  \end{pgfonlayer}

\end{tikzpicture}
}
\vspace{-5mm}
	\caption{A list with head pointers.}
  \label{fig:sll_head_ptr}
\end{center}
\end{wrapfigure}

\emph{Uniting knots.} If two different confluences $\confl$ and $\confl'$ share
an edge, then after folding $\confl$, the resulting edge shares with $\confl'$
two nodes (at least one being a target node), and thus $\confl'$ contains
a~join of $\iograph$. To eliminate this join too, both confluences must be
folded together. A similar reasoning may be repeated with knots in general.
Usefulness of this rule may be illustrated by an example of the set of lists
with head pointers. Without uniting, every list would generate a hierarchy of
knots of the same depth as the length of the list, as illustrated in
Fig.~\ref{fig:sll_head_ptr}. This is clearly impractical since the entire set
could not be represented using finitely many boxes. Rule~\ref{enum:knot_union}
unites all knots into one that contains the entire list, and the set of all
such knots can then be represented by a single FA (containing a loop accepting
the inner nodes of the lists).

\emph{Complexity of knots.} The notion of complexity is introduced to limit the
effect of Rule~\ref{enum:knot_union} of the definition of a knot, which unites
knots that share a sub-edge, and to hopefully make it follow the natural
hierarchical structuring of data structures.
Consider, for instance, the case of singly-linked lists (SLLs) of cyclic
doubly-linked lists (DLLs). In this case, it is natural to first fold the
particular segments of the DLLs (denoted as DLSs below), i.e., to introduce a
box for a single pair of next and prev pointers. This way, one effectively
obtains SLLs of cyclic SLLs. Subsequently, one can fold the cyclic SLLs into a
higher-level box. However, uniting all knots with a common sub-edge would
create knots that contain entire cyclic DLLs (requiring unboundedly many joins
inside the box). The reason is that in addition to the confluences
corresponding to DLSs, there are confluences which traverse the entire cyclic
DLLs and that share sub-edges with all DLSs (this is in particular  the case of
the two circular sequences consisting solely of next and prev pointers respectively).
To avoid the undesirable folding, we exploit the notion of complexity and fold
graphs in successive rounds.
In each round we fold all optimal knots with
the smallest complexity (as described in Sec.~\ref{sec:folding_overview}),
which should correspond to the currently most nested, not yet folded,
sub-structures.
In the previous example, the algorithm starts by folding
DLSs of complexity 2, because the complexity of the confluences in
cyclic DLLs is given by the number of the DLSs they traverse.


\emph{Closure of knots.} The closure is introduced for practical reasons. It
allows one to identify optimal knots by their signatures, which is then used to
simplify automata constructions that implement folding on the level of FA (cf.
Sec.~\ref{sec:folding_overview}).

\emph{Root of an optimal knot.} The requirement for an optimal knot $\knot$ to have a
root is to guarantee that if an io-graph $\iographof \graphprime\gportsprime$
containing a box $B$ representing $\knot$ is accessible, then the
io-graph $\iographof \graphprime\gportsprime\subst{k}{B}$ emerging by
substituting $\knot$ for a sub-edge labelled with $B$ is accessible, and vice
versa.
It is also a necessary condition for the existence of a canonical forest
representation of the knot itself (since one needs to order the cut-points
w.r.t.~the prices of the paths leading to them from the input port of the knot).

\subsection{Folding in the Abstraction Loop}
\label{sec:folding_overview}

\begin{wrapfigure}[8]{r}{41mm}
\vspace{-9mm}
\begin{algorithm}[H]
\dontprintsemicolon
{\it Unfold solitaire boxes}\;
\Repeat{fixpoint}
{
	 {\it Normalise}				\;\nllabel{line:normalise}
	 {\it Abstract}					\;\nllabel{line:abstract}	   
		 	 {\it Fold}						\;   \nllabel{line:fold}
}
\vspace{1mm}
\caption{Abstraction Loop}
\label{alg:learning}
\end{algorithm}
\end{wrapfigure}

In this section, we describe the operation of folding together with the main
abstraction loop of which folding is an integral part. The pseudo-code of the
main abstraction loop is shown in Alg.~\ref{alg:learning}. The algorithm
modifies a set of FA until it reaches a fixpoint. Folding on
line~\ref{line:fold} is a sub-procedure of the algorithm which looks for
substructures of FA that accept optimal knots, and replaces these substructures
by boxes that represent the corresponding optimal knots. The operation of
folding is itself composed of four consecutive steps: \emph{Identifying
indices}, \emph{Splitting}, \emph{Constructing boxes}, and \emph{Applying
boxes}. For space reasons, we give only an overview of the steps of the main
abstraction loop and folding. Details may be found in \cite{techrep}.

\begin{wrapfigure}[3]{r}{26mm}
\vspace{-12.5mm}
\begin{center}
\figswitch{
\hspace{-0mm}\parbox{26mm}{\usetikzlibrary{arrows,backgrounds,fit}
\begin{tikzpicture}[
  node distance=6.6mm,
  thick,
  ->,
  >=stealth'
]
  \tikzstyle{vertex}=[circle,draw,inner sep=0.5mm]

  \node[vertex] (n1)  at (0mm,0cm) { };
  \node[vertex] (n2) [right of=n1] { };
  \node[vertex] (n3) [right of=n2] { };
  \node[vertex] (n4) [right of=n3] { };

  \draw (n1) to[bend left] coordinate (upper1) (n2);
  \draw (n2) to[bend left] coordinate (lower1) (n1);
  \draw (n2) to[bend left] (n3);
  \draw (n3) to[bend left] (n2);
  \draw (n3) to[bend left] coordinate (upper2) (n4);
  \draw (n4) to[bend left] coordinate (lower2) (n3);

  \begin{pgfonlayer}{background}

    \node[draw,rectangle,fill=red!20,draw=black,rounded corners=2mm,inner sep=7.3pt,fit= (n4) (n1) (upper1) (lower1)] (r1) {};

    \node[draw,rectangle,fill=blue!30,draw=black,rounded corners=2mm,inner sep=5.4pt,fit= (n4) (n2) (upper2) (lower2)] (r1) {};

  \node[draw,rectangle,fill=red!20,draw=black,rounded corners=2mm,inner sep=3.5pt,fit= (n4) (n3) (upper2) (lower2)] (r1) {};

  \end{pgfonlayer}

\end{tikzpicture}}\\
}
\vspace{-3mm}
  \caption{DLL.}
  \label{fig:dll_knots}
\end{center}
\end{wrapfigure}


\smallskip\noindent\emph{Unfolding of solitaire boxes.} Folding is in practice
applied on FA that accept partially folded graphs (only some of the optimal
knots are folded). This may lead the algorithm to hierarchically fold data
structures that are not hierarchical, causing the symbolic execution not to
terminate. For example, consider a program that creates a DLL of an arbitrary
length. Whenever a new DLS is attached, the folding algorithm would enclose it
into a box together with the tail which was folded previously. This would lead
to creation of a hierarchical structure of an unbounded depth (see
Fig.~\ref{fig:dll_knots}), which would cause the symbolic execution to never
reach a fixpoint. Intuitively, this is a situation when a repetition of
subgraphs may be expressed by an automaton loop that iterates a box, but it is
instead misinterpreted as a recursive nesting of graphs. This situation may
happen when a newly created box contains another box that cannot be iterated
since it does not appear on a loop (e.g, in Fig.~\ref{fig:dll_knots} there is
always one occurrence of a box encoding a shorter DLL fragment inside a
higher-level box). This issue is addressed in the presented algorithm by first
unfolding all occurrences of boxes that are not iterated by automata loops
before folding is started.


\smallskip\noindent\emph{Normalising.} We define the \emph{index} of a
cut-point $u \in \cpsof\iograph$ as its position in the canonical ordering of
cut-points of $\iograph$, and the \emph{index} of a closed subgraph $\subgraph$
of $\iograph$ as the set of indices of the cut-points in $\sigof\subgraph$. The
folding algorithm expects the input FA $\fa$ to satisfy the property that all
io-graphs of $\langof\fa$ have the same indices of closed knots. The reason is
that folding starts by identifying the index of an optimal knot of an arbitrary
io-graph from $\langof\fa$, and then it creates a box which accepts all closed
subgraphs of the io-graphs from $\iograph$ with the same index. We need a
guarantee that \emph{all} these subgraphs are indeed optimal knots. This
guarantee can be achieved if the io-graphs from $\langof\fa$ have equivalent
interconnections of cut-points, as defined below.

We define the relation $\overlapof {\iograph}\subseteq\nat\times\nat$ between
indices of closed knots of $\iograph$ such that $N\overlapof {\iograph} N'$ iff
there is a closed knot $\knot$ of $\iograph$ with the index $N$ and a closed
knot $\knot'$ with the index $N'$ such that $\knot$ and $\knot'$ have
intersecting sets of sub-edges.  We say that two io-graphs $\iograph$ and
$\iographprime$ are \emph{interconnection equivalent} iff
${\overlapof{\iograph}} = {\overlapof{\iographprime}}$.

\begin{lemma} \label{lemma:same_signatures} Interconnection equivalent
io-graphs have the same indices of optimal knots. \end{lemma}

Interconnection equivalence of all io-graphs in the language of an FA $\fa$ is
achieved by transforming $\fa$ to the \emph{interconnection respecting form}.
This form requires that the language of every TA of the FA consists of
interconnection equivalent trees (when viewing root references and roots as
cut-points with corresponding indices). The transformation is described in
\cite{techrep}. The normalisation step also includes a transformation into the
state uniform and canonicity respecting form.


\smallskip\noindent\emph{Abstraction.} We use abstraction described in
Sec.~\ref{sec:abstraction} that preserves the canonicity respecting form of TA
as well as their state uniformity. It may break interconnection uniformity, in
which case it is followed by another round of normalisation. Abstraction is
included into each round of folding for the reason that it leads to learning
more general boxes. For instance, an FA encoding a cyclic list of one
particular length is first abstracted into an FA encoding a set of cyclic lists
of all lengths, and the entire set is then folded into a single box. 


\smallskip\noindent\emph{Identifying indices.}
For every FA $\fa$ entering this sub-procedure,
we pick an arbitrary io-graph $\iograph\in\langof\fa$, find all its
optimal knots of the smallest possible complexity $n$, and extract their indices.
By Lemma~\ref{lemma:same_signatures} and since $\fa$ is normalised, indices of the optimal knots are the same for all io-graphs in $\langof\fa$.
For every found index, 
the following steps fold all optimal knots with that index at once.
Optimal knots of complexity $n$ do not share sub-edges, the order in which
they are folded is therefore not important.



\smallskip\noindent\emph{Splitting.} For an FA $\fa = (\tas,\fports)$ and an
index $I$ of an optimal knot found in the previous step, splitting transforms
$\fa$ into a (set of) new FA with the same language. The nodes of the borders
of $I$-indexed optimal knots of io-graphs from $\langof\fa$ become roots of
trees of io-forests accepted by the new FA. Let $\src\in I$ be a position in $F$ such that
the $\src$-indexed cut-points of io-graphs from $\langof\fa$ reach all the
other $I$-indexed cut-points. The index $\src$ exists since an optimal knot has
a root. Due to the definition of the closure, the border contains all
$I$-indexed cut-points, with the possible exception of $\src$. The $\src$-th
cut-point may be replaced in the border of the $I$-indexed optimal knot by the
base $\entry$ of the $I$-indexed confluence that is the first one reached from
the $\src$-th cut-point by a straight path. We call $\entry$ the \emph{entry}.
The entry $\entry$ is a root of the optimal knot, and the $\src$-th cut-point
is the only $I$-indexed cut-point that might be outside the knot. If $\entry$
is indeed different from the $\src$-th cut-point, then the $\src$-th tree of
forests accepted by $\fa$ must be split into two trees in the new FA: The
subtree rooted at the entry is replaced by a reference to a new tree. The new
tree then equals the subtree of the original $\src$-th tree rooted at the
entry.

The construction is carried out as follows. We find all states and all of their
rules that accept entry nodes. We denote such states and rules as entry states
and rules. For every entry state $q$, we create a new FA $\splitfa$ which is a
copy of $\fa$ but with the $\src$-th TA $\ta_s$ split to a new $\src$-th TA
$\ta_s'$ and a new $(n+1)$-th TA $\ta_{n+1}$. The TA $\ta_s'$ is obtained
from $\ta_s$ by changing the entry rules of $q$ to accept just a reference to
the new $(n+1)$-th root and by removing entry rules of all other entry states (the
entry states are processed separately in order to preserve possibly different
contexts of entry nodes accepted at different states). The new TA $\ta_{n+1}$
is a copy of $\ta_\src$ but with the only accepting state being $q$. Note that
the construction is justified since due to state uniformity, each node that is
accepted by an entry rule and that does not appear below a node that is also
accepted by an entry rule is an entry node. In the result, the set $J =
(I\setminus\{\src\})\cup\{n+1\}$ contains the positions of the trees of forests
of $\splitfa$ rooted at the nodes of the borders of $I$-indexed optimal knots.




\smallskip\noindent\emph{Constructing boxes.} For every $\splitfa$ and $J$
being the result of splitting $\fa$ according to an index $I$, a box $B_q$ is
constructed from $\splitfa$. We transform TA of $\splitfa$ indexed by the
elements of $J$. The resulting TA will accept the original trees up to that the
roots are stripped from the children that cannot reach a reference to $J$. To
turn these TA into an FA accepting optimal knots with the index $I$, it remains
to order the obtained TA and define port indices, which is described in detail
in \cite{techrep}. Roughly, the input index of the box will be the position $j$
to which we place the modified $(n+1)$-th TA of $\splitfa$ (the one that
accepts trees rooted at the entry). The output indices are the positions of the
TA with indices $J\setminus\{j\}$ in $\splitfa$ which accept trees rooted at
cut-points of the border of the optimal knots.

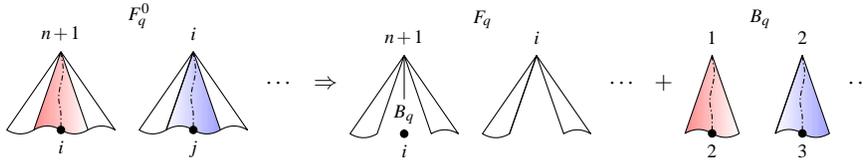
\begin{figure}[t]
  \begin{minipage}{12cm}
	\figswitch{
    \centering
\begin{tikzpicture}[
  edge from parent path=,
  level distance=1.3cm,
  level 1/.style={sibling distance=0.45cm},
  scale=0.8,
  transform shape
]

  \tikzstyle{ref}=[circle,fill=black,inner sep=0.5mm]
  \tikzstyle{dashdot}=[dashed,dash pattern=on 2pt off 1pt on 0.5pt off 1pt]
  \tikzstyle{mysnake}=[decorate,decoration={snake,amplitude=0.3mm,segment length=0.75cm}]

  \node (c1) {}
    child { node (c11) {} }
    child { node (c12) {} }
    child { node (c13) [ref,label={below:$\refinfigof{i}$}] {} }
    child { node (c14) {} }
    child { node (c15) {} };

  \node (c10) [minimum height=0.5cm,node distance=0.3cm,above of=c1] {$n+1$};
  \node (name)[right of=c10,node distance =13mm,yshift=3mm] {$\splitfa$};

  \draw [join=round] (c1.center)
    to (c11.center)
    to [bend right] (c12.center)
    -- cycle;

  \draw [join=round] (c12.center)
    to [bend left](c13.center) 
    to [bend right] (c14.center)
    to (c1.center);

  \draw [join=round] (c14.center)
    to [bend left] (c15.center)
    to (c1.center);

  \begin{pgfonlayer}{background}
  \path [right color=white,left color=red!50,shading=axis] (c12.center)
    to [bend left] (c13.center)
    to [bend right] (c14.center)
    to (c1.center);
  \draw [mysnake,dashdot] (c1.center) -- (c13.center);
  \end{pgfonlayer}

  \node (c2) [node distance=2.2cm,right of=c1] {}
    child { node (c22) {} }
    child { node (c23) {} }
    child { node (c24) [ref,label={below:$\refinfigof{j}$}] {} }
    child { node (c25) {} }
    child { node (c26) {} };

  \node (c20) [minimum height=0.5cm,node distance=0.3cm,above of=c2] {$i$};

  \draw [join=round] (c2.center) -- (c23.center);

  \draw [join=round] (c2.center)
    to (c22.center) 
    to [bend right] (c23.center)
    -- cycle;

  \draw [join=round] (c23.center)
	to [bend left] (c24.center)
    to [bend right] (c25.center)
    to (c2.center);

  \draw [join=round] (c25.center)
    to [bend left] (c26.center) 
    to (c2.center);

  \begin{pgfonlayer}{background}
  \path [left color=white,right color=blue!40,shading=axis] (c23.center)
    to [bend left] (c24.center)
    to [bend right] (c25.center)
    to (c2.center);
  \draw [mysnake,dashdot] (c2.center) -- (c24.center);
  \end{pgfonlayer}

\end{tikzpicture}
    \raisebox{1.0cm}{$\cdots \ \ \ \Rightarrow$}
\begin{tikzpicture}[
  edge from parent path=,
  level distance=1.3cm,
  level 1/.style={sibling distance=0.45cm},
  scale=0.8,
  transform shape
]

  \tikzstyle{ref}=[circle,fill=black,inner sep=0.5mm]
  \tikzstyle{dashdot}=[dashed,dash pattern=on 2pt off 1pt on 0.5pt off 1pt]
  \tikzstyle{mysnake}=[decorate,decoration={snake,amplitude=0.3mm,segment length=0.75cm}]

  \node (c1) {}
    child { node (c11) {} }
    child { node (c12) {} }
    child { node (c13) [ref,label={below:$\refinfigof{i}$}] {} }
    child { node (c14) {} }
    child { node (c15) {} };

  \node (c10) [minimum height=0.5cm,node distance=0.3cm,above of=c1] {$n+1$};
  \node (name)[right of=c10,node distance =13mm,yshift=3mm] {$\fa_q$};

  \draw [join=round] (c1.center)
    to (c11.center)
    to [bend right] (c12.center)
    -- cycle;


  \draw (c1.center) to node [near end,circle,fill=white,inner sep=0.5pt] {$B_q$} (c13.center);

  \draw [join=round] (c1.center)
	to (c14.center)
    to [bend left] (c15.center)
    to (c1.center);


  \node (c2) [node distance=2.2cm,right of=c1] {}
    child { node (c22) {} }
    child { node (c23) {} }
    child { node (c24) {} }
    child { node (c25) {} }
    child { node (c26) {} };

  \node (c20) [minimum height=0.5cm,node distance=0.3cm,above of=c2] {$i$};

  \draw [join=round] (c2.center) -- (c23.center);

  \draw [join=round] (c2.center)
    to (c22.center) 
    to [bend right] (c23.center)
    -- cycle;


  \draw [join=round] (c25.center)
    to [bend left] (c26.center) 
    to (c2.center) -- cycle;


\end{tikzpicture}
    \raisebox{1.0cm}{$\cdots \ \ \ +$}
\begin{tikzpicture}[
  edge from parent path=,
  level distance=1.3cm,
  level 1/.style={sibling distance=0.45cm},
  scale=0.8,
  transform shape
]

  \tikzstyle{ref}=[circle,fill=black,inner sep=0.5mm]
  \tikzstyle{dashdot}=[dashed,dash pattern=on 2pt off 1pt on 0.5pt off 1pt]
  \tikzstyle{mysnake}=[decorate,decoration={snake,amplitude=0.3mm,segment length=0.75cm}]

  \node (c1) {}
    child { node (c12) {} }
    child { node (c13) [ref,label={below:$\refinfigof{2}$}] {} }
    child { node (c14) {} };

  \node (c10) [node distance=0.3cm,inner sep=0pt,above of=c1] {$1$};
  \node (name)[right of=c10,node distance =8mm,yshift=3mm] {$B_q$};


  \draw [join=round] (c12.center)
    to [bend left](c13.center) 
    to [bend right] (c14.center)
    to (c1.center) -- cycle;


  \node (c2) [node distance=1.5cm,right of=c1] {}
    child { node (c23) {} }
    child { node (c24) [ref,label={below:$\refinfigof{3}$}] {} }
    child { node (c25) {} };

  \node (c20) [node distance=0.3cm,inner sep=0pt,above of=c2] {$2$};

  \draw [join=round] (c2.center) -- (c23.center);


  \draw [join=round] (c23.center)
	to [bend left] (c24.center)
    to [bend right] (c25.center)
    to (c2.center) -- cycle;



  \begin{pgfonlayer}{background}

  \path [right color=white,left color=red!50,shading=axis] (c12.center)
    to [bend left] (c13.center)
    to [bend right] (c14.center)
    to (c1.center);
  \draw [mysnake,dashdot] (c1.center) -- (c13.center);

  \path [left color=white,right color=blue!40,shading=axis] (c23.center)
    to [bend left] (c24.center)
    to [bend right] (c25.center)
    to (c2.center);
  \draw [mysnake,dashdot] (c2.center) -- (c24.center);
  \end{pgfonlayer}

\end{tikzpicture}
    \raisebox{1.0cm}{$\cdots$}
    }
  \end{minipage}
  \vspace*{-1mm}
  \caption{Creation of $\fa_q$ and $B_q$ from $\splitfa$. 
The subtrees that contain references $i,j\in J$ are taken into $B_q$, 
and replaced by the $B_q$-labelled sub-edge in $\fa_q$.
}
  \label{fig:slicing}
  \vspace*{-3mm}
\end{figure}


\smallskip\noindent\emph{Applying boxes.} This is the last step of folding\tronly{ (\emph{hallelujah!})}. For every $\splitfa$, $J$, and $B_q$ which are the result
of splitting $\fa$ according to an index $I$, we construct an~FA $\fa_q$ that
accepts graphs of $\fa$ where knots enclosed in $B_q$ are substituted by a
sub-edge with the label $B_q$. It is created from $\splitfa$ by (1) leaving out
the parts of root rules of its TA that were taken into $B_q$, and (2) adding
the rule-term $B_q(r_1,\ldots,r_m)$ to the rule-terms of root rules of the
$(n+1)$-th component of $\splitfa$ (these are rules used to accept the roots of
the optimal knots enclosed in $B_q$). The states $r_1,\ldots,r_m$ are fresh
states that accept root references to the appropriate elements of $J$ (to
connect the borders of knots of $B_q$ correctly to the graphs of $\fa_q$---the
details may be found in \cite{techrep}). The FA $\fa_q$ now accepts graphs
where optimal knots of graphs of $\langof\fa$ with the signature $I$ are hidden
inside $B_q$. Creation of $B_q$ and of its counterpart $F_q$ from $\splitfa$ is
illustrated in Fig.~\ref{fig:slicing} where $i,j,\ldots\in J$.



During the analysis, the discovered boxes must be stored in a database and
tested for equivalence with the newly discovered ones since the alphabets of FA
would otherwise grow with every operation of folding \emph{ad infinitum}. That
is, every discovered box is given a unique name,\tronly{ such as ``DLL'' for
the box from Fig.~\ref{fig:dll},} and whenever a~semantically equivalent box is
folded, the newly created edge-term is labelled by that name. This step offers
an opportunity for introducing another form of acceleration of the symbolic
computation. Namely,~when~a~box~$B$ is found by the procedure described above,
and another box $B'$ with a name $N$ s.t.  $\semof {B'} \subset \semof {B}$ is
already in the database, we associate the name $N$ with $B$ instead of with
$B'$ and restart the analysis
(i.e., start the analysis from the scratch, remembering just the updated database of boxes). 
If, on the other hand, $\semof {B} \subseteq
\semof {B'}$, the folding is performed using the name $N$ of $B'$, thus
overapproximating the semantics of the folded FA. As presented in
Sec.~\ref{sec:experiments}, this variant of the procedure, called \emph{folding
by inclusion}, performs in some difficult cases significantly better than
the former variant, called \emph{folding by equivalence}.

\section{Abstraction}\label{sec:abstraction}
\rem{Could be maybe shortened using a massacre.}

The abstraction we use in our analysis is based on the general techniques
described in the framework of abstract regular (tree) model checking
~\cite{artmc12}. We, in particular, build on the \emph{finite height
abstraction} of TA. It is parameterised by a height $k\in \nat$, and it
collapses TA states $q,q'$ iff they accept trees with the same sets of prefixes
of the height at most $k$ (the prefix of height $k$ of a tree is a subgraph of
the tree which contains all paths from the root of length at most $k$). This
defines an equivalence on states denoted by $\eql_k$. The equivalence $\eql_k$
is further refined to deal with various features special for FA. Namely, it has
to work over tuples of TA and cope with the interconnection of the TA via root
references, with the hierarchical structuring, and with the fact that we use a
\emph{set} of FA instead of a single FA to represent the abstract context at a
particular program location.


\smallskip\noindent\emph{Refinements of $\eql_k$.} First, in order to maintain
the same basic shape of the heap after abstraction (such that no cut-point
would, e.g.,~suddenly appear or disappear), we refine $\eql_k$ by requiring
that equivalent states must have the same spans (as defined in
Sec.~\ref{sec:fa}). When applied on $\eql_1$, which corresponds to equivalence
of data types, this refinement provided enough precision for most of the case
studies presented later on, with the exception of the most difficult ones,
namely programs with skip lists~\cite{skiplists}. To verify these programs, we
needed to further refine the abstraction to distinguish automata states
whenever trees from their languages encode tree components containing a
different number of unique paths to some root reference, but some of these
paths are hidden inside boxes. In particular, two states $q,q'$ can be
equivalent only if for every io-graph $\iograph$ from the graph language of the FA,
for every two nodes $u,v\in\domof\iograph$ accepted by $q$ and $q'$,
respectively, in an accepting run of the corresponding TA, the following holds:
For every $w\in\cpsof\iograph$, both $u$ and $v$ have the same number of
outgoing sub-edges (selectors) in $\semof\iograph$ which start a trace in
$\semof\iograph$ leading to $w$. According to our experiments, this refinement
does not cost almost any performance, and hence we use it by default. 


\smallskip\noindent\emph{Abstraction for Sets of FA.} Our analysis works with
sets of FA. We observed that abstracting individual FA from a set of FA in
isolation is sometimes slow since in each of the FA, the abstraction widens
some selector paths only, and it takes a while until an FA in which all possible
selector paths are widened is obtained. For instance, when analysing a program
that creates binary trees, before reaching a~fixpoint, the symbolic analysis
generates many FA, each of them accepting a subset of binary trees with some of
the branches restricted to a bounded length (e.g., trees with no right
branches, trees with a single right branch of length 1, length 2, etc.). In
such cases, it helps when the abstraction has an opportunity to combine
information from several FA. For instance, consider an FA that encodes binary
trees degenerated to an arbitrarily long left branch, and another FA that
encodes trees degenerated to right branches only. Abstracting these FA in
isolation has no effect. However, if the abstraction is allowed to collapse
states from both of these FA, it can generate an FA accepting all possible
branches. 

Unfortunately, the natural solution to achieve the above, which is to unite FA
before abstraction, cannot be used since FA are not closed under union (uniting
TA component-wise overapproximates the union). However, it is possible to
enrich the automata structure of an FA $\fa$ by TA states and rules of another
one without changing the language of $\fa$, and in this way allow the
abstraction to combine the information from both FA.  In particular, before
abstracting an FA $\fa = (\tas,\fports)$ from a set $S$ of FA, we pre-process
it as follows.  \begin{inparaenum}[(1)] \item We pick automata $\fa' =
(\ta_1'\cdots\ta_n',\fports)\in S$ which are compatible with $\fa$ in that they
have the same number of TA, the same port references, and for each $1\leq i
\leq n$, the root states of $\ta_i'$ have the same spans as the root states
of $\ta_i$. \item For all such $\fa'$ and each $1\leq i \leq n$, we add rules
and states of $\ta_i'$ to $\ta_i$, but we keep the original set of root
states of $\ta_i$. Since we assume that the sets of state of TAs of different
FA are disjoint, the language of $\ta_i$ stays the same, but its structure is
enriched, which helps the abstraction to perform a coarser widening.
\end{inparaenum}

\section{Experimental Results}\label{sec:experiments}

We have implemented the above proposed techniques in the Forester tool and
tested their generality and efficiency on a number of case studies. In the
experiments, we compare two configurations of Forester, and we also compare the
results of Forester with those of Predator \cite{predator11}, which uses a
graph-based memory representation inspired by separation logic with higher-order
list predicates. We do not provide a comparison with Space
Invader~\cite{InvaderCAV08} and SLAyer~\cite{slayer11}, based also on separation
logic with higher-order list predicates, since in our experiments they were
outperformed by Predator.

In the experiments, we considered programs with various types of lists (singly 
and doubly linked, cyclic, nested, with skip pointers), trees, and their
combinations. In the case of skip lists, we had to slightly modify the
algorithms since their original versions use an ordering on the data stored in
the nodes of the lists (which we currently do not support) in order to guarantee
that the search window delimited on some level of skip pointers is not left on
any lower level of the skip pointers. In our modification, we added an
additional explicit end-of-window pointer. We checked the programs for memory
safety only, i.e., we did not check data-dependent properties.

Table~\ref{tab:results} gives running times in seconds (the average of 10
executions) of the tools on our case studies. ``Basic'' stands for Forester with
the abstraction applied on individual FA only and ``SFA'' stands for
Forester with the abstraction for sets of FA. The value T means that the running
time of the tool exceeded 30 minutes, and the value Err means that the tool
reported a spurious error. The names of the examples in the table contain the
name of the data structure manipulated in the program, which is ``SLL'' for
singly linked lists, ``DLL'' for doubly linked lists (the ``C'' prefix denotes
cyclic lists), ``tree'' for binary trees, ``tree+parents'' for trees with parent
pointers. Nested variants of SLL (DLL) are named as ``SLL (DLL) of'' and the
type of the nested structure.  In particular, ``SLL of 0/1 SLLs'' stands for SLL
of a nested SLL of length 0 or 1, and ``SLL of 2CDLLs'' stands for SLL whose
each node is a root of two CDLLs. The ``+head'' flag stands for a list where
each element points to the head of the list and the subscript ``Linux'' denotes
the implementation of lists used in the Linux kernel, which uses type casts and
a restricted pointer arithmetic. The ``DLL+subdata'' stands for a~kind of a~DLL
with data pointers pointing either inside the list nodes or optionally outside of them.
For a ``skip list'', the subscript denotes
the number of skip pointers. In the example ``tree+stack'', a randomly
constructed tree is deleted using a stack, and ``DSW'' stands for the
Deutsch-Schorr-Waite tree traversal (the Lindstrom variant). All experiments
start with a random creation and end with a~disposal of the specified structure
while the indicated procedure (if any) is performed in between. The experiments
were run on a~machine with the Intel i7-2600 (3.40\,GHz) CPU and 16\,GiB of RAM.

\begin{table}[t]
  \vspace*{-1mm}
  \centering
  \footnotesize
  \caption{Results of the experiments}
  \begin{tabular}{|l||r|r|rcl|r|}
    \hline Example                           & basic & SFA & \multicolumn{3}{c|}{boxes}  & Predator   \\ \hline
    \hline SLL (delete)                      & 0.03  & 0.04 &   & &                      & 0.04       \\
           SLL (bubblesort)                  & 0.04  & 0.04 &   & &                      & 0.03       \\
           SLL (mergesort)                   & 0.08  & 0.15 &   & &                      & 0.10       \\
           SLL (insertsort)                  & 0.05  & 0.05 &   & &                      & 0.04       \\
           SLL (reverse)                     & 0.03  & 0.03 &   & &                      & 0.03       \\
           SLL+head                          & 0.05  & 0.05 &   & &                      & 0.03       \\
           SLL of 0/1 SLLs                   & 0.03  & 0.03 &   & &                      & 0.11       \\
           SLL$_{\mathrm{Linux}}$            & 0.03  & 0.03 &   & &                      & 0.03       \\
           SLL of CSLLs                      & 2.07  & 0.73 &  3&/&4                     & 0.12       \\
           SLL of 2CDLLs$_{\mathrm{Linux}}$  & 0.16  & 0.17 & 13&/&5                     & 0.25       \\
    \hline skip list$_2$                     & 0.66  & 0.42 &  -&/&3                     & T          \\
           skip list$_3$                     & T     & 9.14 &  -&/&7                     & T          \\
    \hline
  \end{tabular}
  \begin{tabular}{|l||r|r|rcl|r|}
    \hline Example                           & basic & SFA & \multicolumn{3}{c|}{boxes}  & Predator   \\ \hline
    \hline DLL (reverse)                     & 0.04  & 0.06 &  1&/&1                     & 0.03       \\
           DLL (insert)                      & 0.06  & 0.07 &  1&/&1                     & 0.05       \\
           DLL (insertsort1)                 & 0.35  & 0.40 &  1&/&1                     & 0.11       \\
           DLL (insertsort2)                 & 0.11  & 0.12 &  1&/&1                     & 0.05       \\
           DLL of CDLLs                      & 5.67  & 1.25 &  8&/&7                     & 0.22       \\
           DLL+subdata                       & 0.06  & 0.09 &  -&/&2                     & T          \\
           CDLL                              & 0.03  & 0.03 &  1&/&1                     & 0.03       \\
    \hline tree                              & 0.14  & 0.14 &   & &                      & Err        \\
           tree+parents                      & 0.18  & 0.21 &  2&/&2                     & T          \\
           tree+stack                        & 0.09  & 0.08 &   & &                      & Err        \\
           tree (DSW)                        & 1.74  & 0.40 &   & &                      & Err        \\
           tree of CSLLs                     & 0.32  & 0.42 &  -&/&4                     & Err        \\
    \hline
  \end{tabular}
  \vspace*{-4mm}
  \label{tab:results}
\end{table}


The table further contains the column ``boxes'' where the value ``X/Y'' means
that X manually created boxes were provided to the analysis that did not use
learning while Y boxes were learnt when the box learning procedure was enabled.
The value ``-'' of X means that we did not run the given example with manually
constructed boxes since their construction was too tedious. If user-defined
boxes are given to Forester in advance, the speedup is in most cases negligible,
with the exception of ``DLL of CDLLs'' and ``SLL of CSLLs'', where it is up to 7
times.
In a majority of cases, the learnt boxes were the same as the ones created manually.
However,
in some cases, such as ``SLL of
2CDLLs$_{\mathrm{Linux}}$'', the learning algorithm found a smaller set of more elaborate boxes than those provided manually.

In the experiments, we use folding by inclusion as defined in
Sec.~\ref{sec:folding_overview}. For simpler cases, the performance matched
the performance of folding by equivalence, but for the more difficult examples
it was considerably faster (such as for ``skip list$_2$'' when the time
decreased from 3.82\,s to 0.66\,s), and only when it was used the analysis of
``skip list$_3$'' succeeded.
Further, the implementation folds optimal knots of the complexity $\leq 2$ which is
enough for the considered examples.
Finally, note that the performance of Forester in
the considered experiments is indeed comparable with that of Predator even
though Forester can handle much more general data structures.

\section{Conclusion}\label{sec:connclusion}

We have proposed a new shape analysis using forest automata which---unlike the
previously known approach based on FA---is fully automated. For that purpose, we
have proposed a technique of automatically learning FA called boxes to be used
as alphabet symbols in higher-level FA when describing sets of complex heap
graphs. We have also proposed a way how to efficiently integrate the learning
with the main analysis algorithm. Finally, we have proposed a significant
improvement---both in terms of generality as well as efficiency---of the
abstraction used in the framework. An implementation of the approach in the
Forester tool allowed us to fully-automatically handle programs over quite
complex heap structures, including 2-level and 3-level skip lists, which---to
the best of our knowledge---no other fully-automated verification tool can
handle. At the same time, the efficiency of the analysis is comparable with
other state-of-the-art analysers even though they handle less general classes of
heap structures.


For the future, there are many possible ways how the presented approach can be
further extended. First, one can think of using recursive boxes or forest
automata using hedge automata as their components in order to handle even more
complex data structures (such as mcf trees). Another interesting direction is
that of integrating FA-based heap analysis with some analyses for dealing with
infinite non-pointer data domains (e.g., integers) or parallelism. 

\smallskip\noindent\emph{Acknowledgement.} This work was supported by the Czech
Science Foundation (projects P103/10/0306, 13-37876P), the Czech Ministry of
Education, Youth, and Sports (project MSM 0021630528), the BUT FIT project
FIT-S-12-1, and the EU/Czech IT4Innovations Centre of Excellence project
CZ.1.05/1.1.00/02.0070.


\rem{for a little happier face, redefine notes and todos (in macros...)}

\vfill\eject\appendix




\begin{thebibliography}{10}

\bibitem{slayer11}
J.~Berdine, B.~Cook, and S.~Ishtiaq.
\newblock {Memory Safety for Systems-level Code.}
\newblock In {\em Proc. of CAV'11}, {\em LNCS} 6806, Springer, 2011.

\bibitem{artmc12}
A.~Bouajjani, P.~Habermehl, A.~Rogalewicz, and T.~Vojnar.
\newblock {Abstract Regular (Tree) Model Checking.}
\newblock {\em STTT} 14(2), Springer, 2012.

\bibitem{rival11}
B.-Y.E. Chang, X.~Rival, and G.C.~Necula.
\newblock {Shape Analysis with Structural Invariant Checkers.}
\newblock In {\em Proc. of SAS'07}, {\em LNCS} 4634, Springer, 2007.

\bibitem{predator11}
K.~Dudka, P.~Peringer, and T.~Vojnar.
\newblock {Predator: A~Practical Tool for Checking Manipulation of Dynamic Data
  Structures Using Separation Logic.}
\newblock In {\em Proc. of CAV'11}, {\em LNCS} 6806, 2011.

\bibitem{indPrSynt07}
B.~Guo, N.~Vachharajani, and D.I.~August.
\newblock {Shape Analysis with Inductive Recursion Synthesis.}
\newblock In Proc. of PLDI'07, ACM Press, 2007.

\bibitem{forester11}
P.~Habermehl, L.~Hol{\'i}k, A.~Rogalewicz, J.~{\v S}im{\'a}{\v c}ek, and
  T.~Vojnar.
\newblock {Forest Automata for Verification of Heap Manipulation.}
\newblock In {\em Proc. of CAV'11}, LNCS 6806, Springer, 2011.

\bibitem{techrep}
L.~Hol{\'i}k, O.~Leng\'al, A.~Rogalewicz, J.~{\v S}im{\'a}{\v c}ek, and
  T.~Vojnar.
\newblock {Fully Automated Shape Analysis Based on Forest Automata.}
\newblock Tech.\ rep.\ FIT-TR-2013-01, FIT BUT, 2013.

\bibitem{juggrnaut10}
J.~Heinen, T.~Noll, and S.~Rieger.
\newblock {Juggrnaut: Graph Grammar Abstraction for Unbounded Heap Structures.}
\newblock ENTCS 266, Elsevier, 2010.

\bibitem{overlaid11}
O.~Lee, H.~Yang, and R.~Petersen.
\newblock {Program Analysis for Overlaid Data Structures.}
\newblock In {\em Proc. of CAV'11}, LNCS 6806, Springer, 2011.

\bibitem{thor10}
S.~Magill, M.-H.~Tsai, P.~Lee, and Y.-K.~Tsay.
\newblock {Automatic Numeric Abstractions for Heap-manipulating programs.}
\newblock In {\em Proc. of POPL'10}, ACM Press, 2010.

\bibitem{juggrnaut-learning12}
A.D.~Weinert. 
\newblock {Inferring Heap Abstraction Grammars.}
\newblock BSc thesis, RWTH Aachen, 2012.

\bibitem{InvaderCAV08}
H.~Yang, O.~Lee, J.~Berdine, C.~Calcagno, B.~Cook, D.~Distefano, and P.W.
  O'Hearn.
\newblock {Scalable Shape Analysis for Systems Code.}
\newblock In {\em Proc. of CAV'08}, {\em LNCS} 5123, Springer, 2008.

\bibitem{skiplists}
W.~Pugh.
\newblock  Skip Lists: A Probabilistic Alternative to Balanced Trees.
\newblock  Commun.\ ACM, 33(6): 668--676, ACM, 1990.

\end{thebibliography}
\end{document}